\documentstyle[12pt,epsfig]{article}
\def\beq{\begin{equation}}   \def\eeq{\end{equation}}
\newcommand{\bea}{\begin{eqnarray}}
\newcommand{\eea}{\end{eqnarray}}

\newcommand{\gsim}{\lower.7ex\hbox{$
\;\stackrel{\textstyle>}{\sim}\;$}}
\newcommand{\lsim}{\lower.7ex\hbox{$
\;\stackrel{\textstyle<}{\sim}\;$}}

\newcommand{\ra}{\rightarrow}

\newcommand{\Lam}{\Lambda_{\rm QCD}}

\newcommand{\as}{\alpha_s}

\newcommand{\GeV}{\,\mbox{GeV}}
\newcommand{\MeV}{\,\mbox{MeV}}

\newcommand{\matel}[3]{\langle #1|#2|#3\rangle}
\newcommand{\aver}[1]{\langle #1\rangle}
\newcommand{\state}[1]{|#1\rangle}
\renewcommand{\ra}{\rightarrow}

\begin{document}

\def\lsim{\mathrel{\rlap{\lower3pt\hbox{\hskip0pt$\sim$}}
    \raise1pt\hbox{$<$}}}         %less than or approx. symbol
\def\gsim{\mathrel{\rlap{\lower4pt\hbox{\hskip1pt$\sim$}}
    \raise1pt\hbox{$>$}}}         %greater than or approx. symbol

\begin{titlepage}
\renewcommand{\thefootnote}{\fnsymbol{footnote}}

\begin{flushright}
UND-HEP-98-BIG\hspace*{.2em}03\\
TECHNION-PH/98-01\\
hep-ph/9805488\\
\end{flushright}
\vspace{.3cm}
\begin{center} \Large
{\bf Four-fermion Heavy Quark Operators and Light Current Amplitudes in
Heavy Flavor Hadrons}
\end{center}
\vspace*{.3cm}
\begin{center} {\Large
D. Pirjol $^a$ \hspace*{.02em} {\large and} \hspace*{.02em}
N. Uraltsev $^{b,c,d}$
}
\vspace*{.7cm}\\

{\small
$^a$ {\it Department of Physics, Israel Institute of Technology,
Technion, Haifa 32000, Israel\,\footnote{Address after September~1:
Newman Laboratory of Nuclear Studies, Cornell University, Ithaca, NY 14853.
}}\\
$^b$ {\it Department of Physics,
University of Notre Dame du Lac, Notre Dame, IN 46556} \\
$^c$ {\it  Theoretical Physics Institute, University of Minnesota,
Minneapolis, MN 55455}\\
$^d$ {\it St.\,Petersburg Nuclear Physics Institute,
Gatchina, St.\,Petersburg 188350, Russia\footnote{Permanent address}
}\\
}
\vspace*{1.2cm}

{\Large{\bf Abstract}}
\end{center}

\vspace*{.2cm}

We introduce and study the properties of the
``color-straight'' four-quark operators containing heavy and light
quark fields. They are of the form
$(\bar b\Gamma_b b)(\bar q\Gamma_q q)$
where both brackets are color singlets.
Their expectation values include the bulk of the nonfactorizable
contributions to the nonleptonic decay widths of heavy hadrons.
The expectation values of the color-straight operators
in the heavy hadrons are related to the momentum integrals of the
elastic light-quark formfactors of the respective heavy hadron.
We calculate the asymptotic behavior of the light-current formfactors
of heavy hadrons and show that the actual decrease is
$1/(q^2)^{\frac32}$ rather than $1/q^4$. The two-loop
hybrid anomalous dimensions of the four-quark operators and their
mixing (absent in the first loop) are obtained.
Using plausible models for the elastic
formfactors, we estimate the expectation values of the color-straight
operators in the heavy mesons and baryons. Improved estimates will
be possible in the future with new data on the radiative decays of
heavy hadrons. We give the Wilson coefficients of the four-fermion
operators in the $1/m_b$ expansion of the inclusive widths and discuss
the numerical predictions. Estimates of the nonfactorizable
expectation values are given.
\vfill
\noindent
\end{titlepage}

\addtocounter{footnote}{-2}

\newpage

\tableofcontents

\section{Introduction}

The heavy quark expansion proved to be useful in describing decay
properties of beauty hadrons. At the level of nonperturbative effects a
number of local heavy quark operators of increasing dimension appears
whose expectation values in the heavy flavor hadrons determine the
importance of preasymptotic effects. The first nontrivial operators,
chromomagnetic $O_G= \bar{b} \frac{i}{2} \sigma_{\mu\nu} G_{\mu\nu} b$
and the kinetic operator $O_\pi= \bar{b} (i\vec{D})^2 b$ have $D=5$.
The expectation value of $O_G$ is known directly from the masses of
beauty hadrons. The expectation value $\mu_\pi^2$ of $O_\pi$ is not yet
known definitely, although a certain progress has been achieved over
the last few years in evaluating it for $B$ mesons.

More operators appear at $D=6$, in particular, four-fermion operators
$\bar{b} \Gamma b \, \bar{q} \Gamma' q$ where $q$ are light quarks and
$\Gamma$, $\Gamma'$ denote various Lorentz and color structures. In the
inclusive widths of heavy hadrons such expectation values govern
$1/m_b^3$ corrections. Their effect is still significant, especially
due to specific accidental suppression of the impact of the leading
$D=5$ operators.

Unfortunately, the expectation values of the four-fermion operators up
to now remain rather uncertain. Since the mid-80s \cite{vsold}, the
vacuum factorization approximation has been used to estimate the
mesonic matrix elements which then appear proportional to $f_B^2$. Such
factorizable terms are absent in baryons, and a number of simple
constituent quark model estimates have been employed.  The validity of
the assumptions implemented in such analyses is not clear, however. As a
result, the expectation values where the factorizable contributions are
absent or suppressed, remain uncertain.

On the other hand, the problem of a more reliable evaluation of the
relevant four-fermion expectation values recently attracted a renewed
attention since the lifetime ratios of the different beauty hadrons
have been accurately measured. While data and predictions of the meson
lifetimes are non-trivially consistent, the small experimental ratio
$\tau_{\Lambda_b}/\tau_{B_d} = 0.78 \pm 0.07$ \cite{data}, if taken
literally, seems to be in a conflict with the expectations based
on the $1/m_b$ expansion.

In the framework of nonrelativistic quark description the four-fermion
expectation values are all expressed via the wavefunction density at
origin $|\Psi(0)|^2$ (for mesons) or the diquark density
$\int d^3y |\Psi(0,y)|^2$ (for baryons). All expectation values differ
then by only simple color and spin factors \cite{vsold}. For example,
in B mesons one has
$$
\frac{1}{2M_B}\matel{B^-}{(\bar{b}b)(\bar{u} u)}{B^-} \;=\;
|\Psi(0)|^2 \;,
$$
\beq
\frac{1}{2M_B}\matel{B^-}{(\bar{b}i\gamma_5 u)(\bar{u}i\gamma_5 b)}{B^-}
\;=\; N_c |\Psi(0)|^2
\label{5}
\eeq
(color indices are contracted inside each bracket), and for baryons
$$
\frac{1}{2M_{\Lambda_b}}\matel{\Lambda_b}{(\bar{b}b)(\bar{u}u)}
{\Lambda_b} \;=\;\int d^3y |\Psi(0,y)|^2  \;,
$$
\beq
\frac{1}{2M_B}\matel{\Lambda_b}{(\bar{b} u)(\bar{u} b)}
{\Lambda_b}\;=\; \frac{1}{2} \; \int d^3y |\Psi(0,y)|^2 \;,
\label{6}
\eeq
etc.

In actual QCD this simple picture does not hold, and the naive Quantum
Mechanics (QM) relations between different expectation values are
generally violated. Moreover, the notion itself of the nonrelativistic
wavefunction used in the potential description, becomes ambiguous. Even
in the perturbative domain the expectation values become
scale-dependent, and the renormalization is in general different for
different operators. This manifestly goes beyond the potential
description, even extended for the price of introducing various
light-quark spin wavefunctions in an attempt to account for the
relativistic bispinor nature of the light quark fields.

In this paper we note that there exists nevertheless a natural
generalization of the notion of the wavefunction density, in particular
at origin (the origin is defined as the position of the heavy quark).
It is associated with the expectation values of those four-fermion
operators for which
the $\bar{b} \Gamma b$ bracket is a color singlet. The
color flow for such operators is not disturbed, and we call them
``color-straight'' operators.
Their expectation values in the heavy quark limit $m_b\ra \infty$ are
related to the observable transition amplitudes.
This fact
suggests that they are better candidates for the
operator basis used to parametrize hadronic expectation values in
various applications. Moreover, they are more suitable also for
applying general bounds of the type discussed in \cite{boost}. Such
QM-type inequalities can be formulated more rigorously for these
operators in full QCD.

Knowledge of the light-quark current elastic formfactors of heavy
hadrons would allow one to determine the color-straight expectation
values. Unfortunately, they are practically unknown yet.
Nevertheless, employing reasonable assumptions about their $q^2$
dependence allows more definite estimates of the expectation values. As
the most conservative attitude, they can be viewed as educated
dimensional analysis, with the added bonus of being free of ambiguities
related to {\it ad hoc} powers of $2\pi$ inherent in various naive
dimensional estimates. Such numerically significant uncertainties often
cause controversy in the resulting expectations leading sometimes to
rather surprising phenomenological conclusions. We also think that the
derived relations can be used for an alternative, simple evaluation of
the color-straight expectation values in the lattice heavy quark
simulations.

\section{Color-straight operators and light current amplitudes}

Our main object of interest is the expectation values of the
color-straight operators of the generic type
\beq
\bar{b}_i\Gamma_b b^i\, \bar{q}_j \Gamma_q q^j
\label{9}
\eeq
where $\Gamma_b$, $\Gamma_q$ are arbitrary matrices contracting Lorentz
indices ($\Gamma_q$ can be also a matrix in the light flavor space),
and $i, j$ are color indices. We will consider the heavy quark limit
$m_b \ra \infty$ assuming that the normalization point $\mu$ of the
operators or currents is set much smaller than $m_b$. In this case
there are two nonvanishing types of operators transforming under
rotations of the heavy quark spin as spin-singlet and spin-triplet,
respectively. The corresponding Dirac structure on the heavy side is
$\Gamma_b=1$ and $\Gamma_b = \vec{\gamma}\gamma_5 = \vec\sigma\,$:
\beq
O_{\mbox{\tiny s-s}}\;=\; (\bar{b}b)\, (\bar{q} \Gamma_q q)\;, \qquad
O_{\mbox{\tiny s-tr}}\;=\; (\bar{b}\sigma_k b)\, (\bar{q} \Gamma_q q)\;
\label{11}
\eeq
All possible $\Gamma_b$-structures are reduced to these operators. In
our discussion we always assume that the heavy quark is at rest, $v_\mu
= (1, \vec{0})$, and $v_\mu$ denotes the velocity of the $b$-hadron
$H_b$. Since in the heavy quark limit the $b$ quark spin decouples, we
start for simplicity from considering the spin-singlet operators
$O_{\mbox{\tiny s-s}}$. The straightforward generalization for
$O_{\mbox{\tiny s-tr}}$ will be formulated later.

If a heavy meson were a two-body QM system where, additionally, the
light quark is nonrelativistic as well, the expectation values of
$O_{i}$ measure the meson wavefunction at origin, see the
first of Eqs.~(\ref{5}), and likewise for other matrices $\Gamma_q$ for
which different spin wavefunctions $\Psi(x)$ can enter. In the momentum
representation
\beq
\Psi(0)\;=\; \int\; \frac{{\rm d}^3\vec{p}}{(2\pi)^3}\:\Psi(\vec{p}\,)
\label{14}
\eeq
(we use the normalization where
$\int\, {\rm d}^3\vec{p}/(2\pi)^3\,|\Psi(\vec{p}\,)|^2=1$).

On the other hand, in such a nonrelativistic system the Fourier
transform of the light quark density distribution measures the elastic
transition amplitude (formfactor) of the meson associated with the
scattering on the light quark:
\beq
{\cal F}(\vec{q}\,)\;\equiv \;
\frac{1}{2M_B} \matel{B(\vec{q})}{\bar{q}q(0)}{B(0)}\;=\;
\int\; {\rm d}^3\vec{x}\:\Psi(\vec{x}) \Psi^*(\vec{x})
{\rm e}\,^{-i\vec{q}\vec{x}}\;.
\label{16}
\eeq
The following relation then obviously holds:
\beq
\int\; \frac{{\rm d}^3\vec{q}}{(2\pi)^3}\:  {\cal F}(\vec{q}\,)\; = \;
|\Psi(0)|^2 \;=\; \frac{1}{2M_B}  \matel{B}{(\bar{b}b)(\bar{q}q)(0)}{B}\;.
\label{17}
\eeq
Integrating the transition amplitude over all $\vec{q}$ yields the
local four-fermion expectation value we are interested in. Since we
study a transition induced by scattering on the light quark, the scale
of the transferred momentum is the typical bound-state momentum and is
much smaller than $m_b$.

In actual QCD the simple nonrelativistic picture does not apply. The
light quark is certainly relativistic. Additionally, a two-body
potential description (generally, any fixed-parton wavefunction) can
only be approximately correct, with {\it a priori} unknown accuracy.

It appears, however, that in spite of the fact that neither
Eqs.~(\ref{5}) nor (\ref{16}) can be rigorously written in QCD, the
final relation between the momentum integral of the (elastic)
transition amplitudes and the color-straight expectation values holds
exactly, up to corrections vanishing when $m_b \ra \infty$. It is not
difficult to see, for example, that proceeding from a two-body
nonrelativistic meson to a three-body nonrelativistic baryon does not
modify the relation. We do not illustrate it here, and instead give a
general field-theoretic proof.

Let us start with the operator $O_{\rm s-s} = \bar{b}b\, \bar{q}\Gamma
q$ and consider the corresponding light quark current and its
transition amplitude:
\beq
J_\Gamma(x)\; = \; \bar{q} \Gamma q(x)\:; \qquad
\frac{1}{2M_{H_b}}\matel{\tilde{H}_b(\vec{q}\,)}{J_\Gamma(0)}{H_b(0)}\;=\;
{\cal A}_\Gamma(\vec{q}\,)\;\;.
\label{21}
\eeq
The current does not need to be scalar; any particular component can
even be considered separately. Likewise, the transition amplitude may
not be a true scalar. The initial and final states may differ. The
following relation holds:
\beq
\frac{1}{2M_{H_b}}\matel{\tilde{H}_b(0)}{\bar{b}b \,\bar{q} \Gamma q(0)}
{H_b(0)}\;=\;
\int\; \frac{{\rm d}^3\vec{q}}{(2\pi)^3}\:  {\cal A}_\Gamma (\vec{q})
\; .
\label{22}
\eeq

To prove this relation, we write explicitly the state
$\tilde{H}_b(\vec{q}\,)$ with non-zero momentum as a result of the
Lorentz boost from rest to the velocity $\vec{v}=
\vec{q}/M_{\tilde{H}_b}$:
\beq
\state{\tilde{H}_b(\vec{q}\,)}\;=\;
U\left[L\left(\frac{\vec q}{M_{\tilde{H}_b}}\right)\right]
\state{\tilde{H}_b(0)}\;,
\label{24}
\eeq
where $U\left[L\left(\frac{\vec q}{M_{\tilde{H}_b}}\right)\right]$
is the corresponding Lorentz boost unitary
operator. This operator is given by \cite{wein}
\beq U[L(\vec v)]\; =\;
{\rm e}\,^{ -i\vec n\cdot \vec K \theta}
\qquad \qquad
\sinh{\theta}=|\vec v|, \qquad  \vec n=\frac{\vec v}{|\vec v|}\; ;
\label{26}
\eeq
the boost generators $\vec{K}$ can be expressed in terms of the
symmetric energy-momentum tensor $T_{\mu\nu}$:
\beq
K^i\;=\; \int \;{\rm d}^3\vec{x}\:\left(x^i T^{00}-x^0 T^{0i} \right)
\label{27}
\eeq
($x^0$ is fixed in Eq.~(\ref{24}) and can be put to zero).
Since $\vec{q}$ does not scale with $m_b$, we actually need to retain
only the linear in $\vec{v}$ terms, which leads to simplifications. For
example, the polarization degrees of freedom of $\tilde{H}_b$ (if any)
do not change at the boost.

The whole energy-momentum tensor consists of two parts:
\beq
T_{\mu\nu}\;=\;  T_{\mu\nu}^{\rm light} + T_{\mu\nu}^{\rm heavy} \;=\;
T_{\mu\nu}^{\rm light} + \frac{1}{4}
\bar{b}[\gamma_\mu (i\vec D)_\nu + \gamma_\nu (i\vec D)_\mu -
\stackrel{\leftarrow}{(iD)}_\nu\gamma_\mu -
\stackrel{\leftarrow}{(iD)}_\mu\gamma_\nu] b\, ,
\label{29}
\eeq
where $T_{\mu\nu}^{\rm light}$ is the usual QCD energy-momentum tensor
including only light fields; it is free of the large parameter $m_b$.
In the heavy quark limit we need to retain only the part of
$T_{\mu\nu}$ which is proportional to $m_b$:
\beq
K^i\;=\; \int\: {\rm d}^3x\; x^i T^{00}(x)\;=\;
m_b \int\: {\rm d}^3x\; x^i \, \bar{b}b(x) \;+\;
{\cal O}\left(m_b^0\right)\;.
\label{30}
\eeq
Here we have used the equations of motion for the $b$ field. The
anomalous terms are included in the last term (see, e.g.,
\cite{optical}, Sect.~II). Therefore, we arrive at
\beq
\matel{\tilde{H}_b(\vec{q})}{J_\Gamma(0)}{H_b(0)}\;=\;
\matel{\tilde{H}_b(0)}{\;
{\rm e}\,^{i \int\, {\rm d}^3z \, (\vec{q}\vec{z})\, \bar{b}b(\vec{z})}
\:J_\Gamma(0)\;}{H_b(0)}\;.
\label{32}
\eeq

The heavy quark limit leads to further simplifications: the number of
heavy quarks becomes fixed and $b$ itself becomes static. Then in the
single-$b$ sector the following identity holds:
\beq
{\rm e}\,^{i \int\, {\rm d}^3z \, f(\vec{z}\,)\bar{b}b(\vec{z}\,) }
\;=\;
\int\; {\rm d}^3z \:{\rm e}\,^{i f(\vec{z}\,)}\, \bar{b}b(z) \;.
\label{34}
\eeq
Indeed, in the single-$b$ sector any product of the static $b$ quark
bilinears is very simple:
\beq
(\bar{b} \sigma_1 b)(\vec{z}_1) ...(\bar{b} \sigma_n b)(\vec{z}_n) \;=\;
\delta^3(\vec{z}_1-\vec{z}_n)... \delta^3(\vec{z}_{n-1}-\vec{z}_n)\,
\bar{b} \sigma_1... \sigma_n b(z_n)\; \vert_{{\rm single}\: b}
\label{36}
\eeq
($\sigma_k$ are arbitrary spin matrices). Using this, we obtain
$$
{\rm e}\,^{i \int\, {\rm d}^3z \, f(\vec{z}\,)\bar{b}b(\vec{z}\,) }
\;=\;
\sum_{n=0}^\infty \; \frac{i^n}{n!}\,
\int\; {\rm d}^3z_1...{\rm d}^3z_n  \; f(\vec{z}_1)...f(\vec{z}_n) \:
\bar{b}b(\vec{z}_1)...\bar{b}b(\vec{z}_n) \;=
$$
\beq
\sum_{n=0}^\infty \; \frac{i^n}{n!}\,
\int\; {\rm d}^3z  \; f^n(\vec{z}\,) \bar{b}b(\vec{z}\,) \;=\;
\int\; {\rm d}^3z\; {\rm e}\,^{i f(\vec{z}\,)}\, \bar{b}b(\vec{z}\,) \;.
\label{37}
\eeq
Taking $f(\vec{z})= \vec{q}\vec{z}$ we rewrite Eq.~(\ref{32}) in the
desired form:
\beq
\matel{\tilde{H}_b(\vec{q}\,)}{J_\Gamma(0)}{H_b(0)}\;=\;
\matel{\tilde{H}_b(0)}{\;
\int\; {\rm d}^3z \: {\rm e}\,^{i \vec{q}\vec{z}} \, \bar{b}b(\vec{z})
\:J_\Gamma(0)\;}{H_b(0)}\;.
\label{40}
\eeq

Eq.~(\ref{40}) provides the discussed quantum field-theory
generalization of the notion of the light-quark density $\bar{q}
\Gamma q$ at arbitrary separation; one can define, for example,
\beq
|\Psi_\Gamma(x)|^2_{H_b}\;=\;
\int\; \frac{{\rm d}^3\vec{q}}{(2\pi)^3} \: {\rm e}\,^{i \vec{q}\vec{x}}
\frac{1}{2M_{H_b}} \matel{H_b(\vec{q}\,)}{J_\Gamma(0)}{H_b(0)}\;.
\label{41}
\eeq

In what follows we are interested in the local heavy quark operators,
that is when the light field operators enter at the same point as the
$b$ quark field. It is these operators that appear in the heavy quark
expansion. Integrating Eq.~(\ref{40}) over $\vec{q}$ we get
\beq
\matel{\tilde{H}_b(0)}{\bar{b} b(0) J_\Gamma(0)}{H_b(0)}\; = \;
\int\; \frac{{\rm d}^3\vec{q}}{(2\pi)^3} \:
\matel{\tilde{H}_b(\vec{q}\,)}{J_\Gamma(0)}{H_b(0)}\;.
\label{44}
\eeq

This is our master equation. We see that, in principle, it is even more
general than was stated earlier: $J_\Gamma(0)$ can be arbitrary
gauge-invariant operator composed of the light fields, and not
necessarily a light-quark bilinear. Besides, this relation holds not
only for the truly forward transition matrix elements. The initial and
final state hadrons can be different. Generally, they can even have
different momenta; however, it must be assumed that these momenta are
small compared to $m_b$ -- say, of the typical light hadron mass scale.
Since this equation involves the integration over all transferred
momenta, varying the relative momentum of the final and initial hadrons
have no effect whatsoever, as it should be.

Informative relations emerge, on the other hand, if we vary the heavy
flavor state $\state{\tilde{H}_b}$ (or $\state{H_b}$) within
the corresponding heavy-spin multiplet. Since the $b$-quark
spin decouples, this yields similar relations for the color-straight
spin-triplet operators containing $\bar{b} \vec{\sigma} b$, that is,
with the axial-vector $b$-quark current. In particular,
\beq
\matel{\tilde{H}_b(0)}{\bar{b} \sigma_k b(0) J_\Gamma(0)}{H_b(0)}\; =
\; \int\; \frac{{\rm d}^3 \vec{q}}{(2\pi)^3} \:
\matel{S_k\, \tilde{H}_b(\vec{q}\,)}{J_\Gamma(0)}{H_b(0)}\;,
\label{46}
\eeq
where $\vec{S}/2$ is the $b$-quark spin operator. Formally one obtains
this using, for example, the representation
$$
\state{S_k\, \tilde{H}_b} \; = \; \int\; {\rm d}^3\vec{x} \:
\bar{b} \sigma_k b(x)\, \state{\tilde{H}_b}
$$
and applying relations (\ref{34}), (\ref{36}) generalized to include
the $b$-quark spin matrices. Alternatively, it follows merely from the
heavy-spin symmetry relation between matrix elements of the operators
$\bar{b}bJ_\Gamma(x)$ and $\bar{b} \sigma_k b J_\Gamma(x)\,$.

It is worth to give a less rigorous but a transparent QM derivation
of the master equation Eq.~(\ref{44}). Let us represent the expectation
value of the color-straight operator $\bar{b}sb\, J_\Gamma (0)$  ($s$
is either the unit or a spin matrix) by the sum over possible
intermediate states:  \beq \matel{\tilde{H}_b}{\bar{b} s b(0)\,
J_\Gamma(0)}{H_b}\; = \; \sum_n \int\; \frac{{\rm d}^3\vec{q}}{(2\pi)^3
2E_n} \:  \matel{\tilde{H}_b}{\bar{b} s b(0)}{n(\vec{q}\,)}
\matel{n(\vec{q}\,)}{J_\Gamma(0)}{H_b}\;.
\label{qmder}
\eeq
The states $\state{n(\vec{q}\,)} $ are hadrons with a single $b$ quark.
In the effective theory the integral over momenta must
converge at a hadronic scale which is much smaller than $m_b$. Then
only the elastic transition (i.e., where $\tilde{H}_b$ and $\state{n}$
belong to the same hyperfine multiplet differing, at most, by the
heavy quark spin alignment if $s$ is not a unit matrix) survive in the
sum:  all excited transition amplitudes generated by the heavy quark
current $\bar{b} s b(0)$ are either proportional to $1/m$, or to
velocity $\vec{v}\simeq \vec{q}/m_b$ of the heavy hadron state
$\state{n(\vec{q}\,)}$ \cite{vshqs}. Moreover, since $\vec{v} \ra 0$ the
elastic amplitude is unity up to corrections $\sim q^2/m_b^2$ we
neglect. Thus, Eq.~(\ref{44}) is reproduced.

\thispagestyle{plain}
\begin{figure}[hhh]
 \begin{center}
 \mbox{\epsfig{file=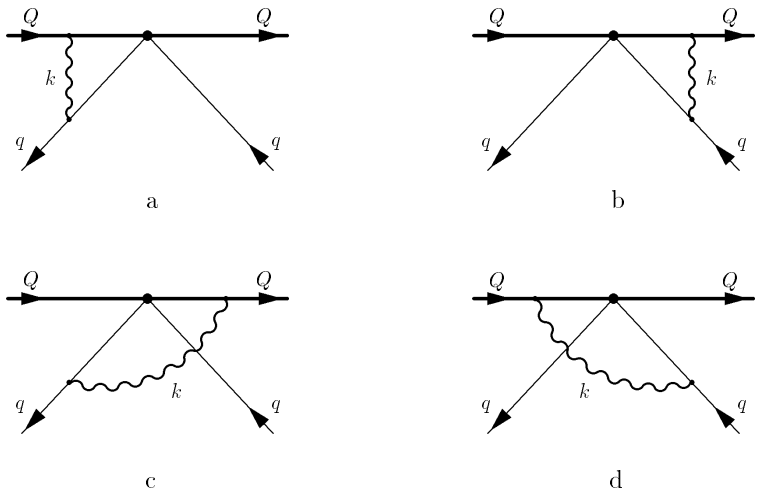,width=10cm}}
 \end{center}
 \caption{
Diagrams for the renormalization of the four-fermion operators.}
\label{fig1}
\end{figure}

Let us illustrate the validity of relation (\ref{44}) diagrammatically,
in respect to the perturbative corrections. Relevant order-$\as$
corrections to the expectation value of the four-fermion operator are
drawn in Figs.~1 whereas Figs.~2 show the corrections to the
formfactor. The gluon exchanges involving only light quarks merely
renormalize the current in question, and we do not consider them.
The corrections dressing the heavy-quark part vanish due to
conservation of the $b$-quark current (we consider gluon momenta much
smaller than $m_b$).

\thispagestyle{plain}
\begin{figure}[hhh]
 \begin{center}
 \mbox{\epsfig{file=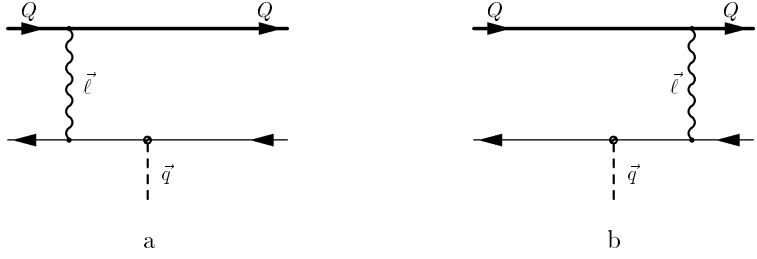,width=10cm}}
 \end{center}
 \caption{
Diagrams contributing to the light quark formfactor of a heavy hadron.}
\label{fig2}
\end{figure}

In the nonrelativistic approximation for the light quark the
``crossed'' diagrams are suppressed, and the remaining diagrams
Figs.~1a and b have obvious counterparts in the corresponding diagrams
in Fig.~2. Going beyond a simple potential approximation (e.g., at $k^2
\gg m_q^2$), however brings in diagrams
Figs.~1\,c,d as well. In fact, one should
keep in mind that in Eq.~(\ref{44}) the integration of the formfactor
is performed only over the spacelike components of $\vec{q}$. This
fixes the spacelike separation of the $\bar{b}b$ and
$\bar{q}\Gamma q$ currents to be zero, however {\it per se} does not
specify the timelike separation of the vertices which is actually
determined by the heavy quark propagators. In reality, a single diagram
Fig.~2a corresponds to the sum of diagrams a and c in Fig.~1, and
likewise with diagrams b. In the coordinate representation, the heavy
quark propagator in Fig.~1a is $\vartheta(-x_0)$ and in Fig.~1c it is
$\vartheta(x_0)$ thus yielding unity in the sum (unity means absence of
any propagation, as in Fig.~2).

Let us illustrate it in the usual momentum representation. Denoting the
gluon momentum in Fig.~1 by $k$, we keep the spacelike components
of $k$ fixed and consider the integral over $\omega \equiv k_0$. The
diagrams a and c are given, respectively, by
\beq
\frac{1}{-\omega-i\epsilon} \cdot {\cal A} \left(\vec{k}, \omega\right)
\qquad\qquad  \mbox{ and } \qquad \qquad
\frac{1}{\omega-i\epsilon} \cdot {\cal A} \left(\vec{k}, \omega\right)
\;,
\label{52}
\eeq
where ${\cal A} \left(\vec{k}, \omega\right)$ generically denotes the
`light' part of the diagram (including the gluon propagator). Since
$$
\frac{1}{-\omega -i\epsilon}\;+\; \frac{1}{\omega-i\epsilon}\;=
\; 2\pi i \delta(\omega)\;,
$$
the integration $\frac{{\rm d}k_0}{2\pi i}$ of the sum of
Figs.~1a and
1c amounts merely to setting $k_0=0$ in the rest of the diagram. (This
is a special case of the more general relations given in Appendix~1.)
Then
it exactly coincides with Fig.~2a if $\vec{k}$ is identified with the
gluon momentum $\vec{l}$ in the latter. Although the gluon momentum
transfer $\vec{l}$ is not generally equal to $\vec{q}$ but can differ
by a primordial momentum in the bound state, integration over all
$\vec{q}$ is equivalent to integration over ${\rm d}^3\vec{l}$.
Similarly, the sum of the diagrams in Figs.~1b and 1d yields the
integral of Fig.~2b over $\vec{q}\,$.

It is clear that this proof is generalized for an arbitrary number of
gluon exchanges between the `light' and `heavy' parts of the diagrams,
or the case of the axial $b$-quark current (see Appendix~1). It is
imperative, however, that the $b$ quark current is color-singlet.

\section{Applications}

We now turn to some applications of the relations (\ref{44}),
(\ref{46}).

\subsection{Perturbative renormalization of the color-straight
operators}

In general, the composite heavy-quark operators depend on the
renormalization point $\mu$ which is assumed to satisfy the `hybrid'
hierarchy condition $\Lam \ll \mu \ll m_b$. The most interesting is the
logarithmic renormalization. This `hybrid' renormalization was first
considered in \cite{fb,vslog,vsku} where the one-loop hybrid anomalous
dimensions were calculated for the quark bilinears and four-fermion
operators.

In the expressions of the matrix elements of the color-straight
operators $\bar{b} (\sigma_k)b \bar{q} \Gamma q$ via the integral of the
transition matrix element of the light quarks current, the
normalization-point dependence can appear in two ways: first, as a
$\mu$-dependence of the light-quark current itself. This is a
usual, `ultraviolet' renormalization since $\mu$ is an ultraviolet
cutoff in respect to the light degrees of freedom. The second way the
dependence on the UV cutoff can enter is via the divergence of the
integral over the momentum of the final state. Indeed, in the
effective theory with the cutoff $\mu$ the perturbative states with
momenta above $\mu$ are absent, while the formfactors with $|\vec{q}\,|
\ll \mu$ coincide with those in full QCD. Therefore, if in full QCD the
integral of the amplitude in Eqs.~(\ref{44},\ref{46}) does not converge
at $\vec{q} \sim \Lam$ but has a $\log$ behavior in the hybrid domain,
this leads to the logarithmic dependence of the matrix element on
$\mu$.

In practice we are interested in vector or axial currents of light
quarks. They are conserved and their anomalous dimensions vanish (for
the flavor-singlet axial-vector current there is an anomalous dimension
in higher orders in $\as$ related to the axial triangle anomaly).
Therefore we will phrase our discussion neglecting this type of
renormalization.

The asymptotics of the actual light quark current formfactors of
the heavy flavor hadrons is given by the perturbative diagrams where
hard gluons transfer the high momentum from
the light quark to the heavy
one. The tree-level order-$\as$ diagrams are shown in Figs.~2\,a,b. By
virtue of the relations Eqs.~(\ref{44}), (\ref{46}) they determine {\it
one-loop} renormalization of the four-fermion operators. It is easy to
see that these diagrams yield amplitudes fading out at least as
$1/\vec{q}^{\,4}$ (the odd powers of $\vec{q}$ do not contribute to the
integral).\footnote{Similar quark counting rules in heavy mesons for
$|\vec{q}\,| \ll m_b$ have been applied, e.g., in \cite{luty}.} In
principle, depending on the particular form of
$\Gamma$, the asymptotics of these diagrams may have the
$1/|\vec{q}\,|^{3}$ term -- it is given by
\beq
\frac{4\pi\as}{\vec{q}^{\,4}}\,
\frac{1}{2M_{H_b}}
\matel{\tilde{H}_b(0)}{\bar{b} t^a b \:
\bar{q}\left(\gamma_0 \!\not \!q \,\Gamma- \Gamma
q\!\!\!\!\!\not \,\gamma_0 \right) t^a q(0)\,}{H_b(0)}\;.
\label{50}
\eeq
($\bar{b} t^a b \; \ra \; \bar{b} \vec{\sigma} t^a b$ for
the spin-flip
transitions). The matrix element may not vanish for beauty hadrons with
nonvanishing spin of light degrees of freedom (let us recall that
$q_0=0$). However, in this matrix element both hadrons are at rest,
therefore any such $1/|\vec{q}\,|^{3}$ term vanishes upon integrating
over the direction of $\vec{q}$.
The fact of vanishing of the
leading-order hybrid anomalous dimension for the operators of the form
$(\bar{b}b)\, (\bar{q} \gamma_\mu(\gamma_5) q)$ was noted in
\cite{vslog} already in the mid 80's as a result of simple calculations
of the one-loop diagrams. Our relation gives it an alternative
interpretation.

\thispagestyle{plain}
\begin{figure}[hhh]
 \begin{center}
 \mbox{\epsfig{file=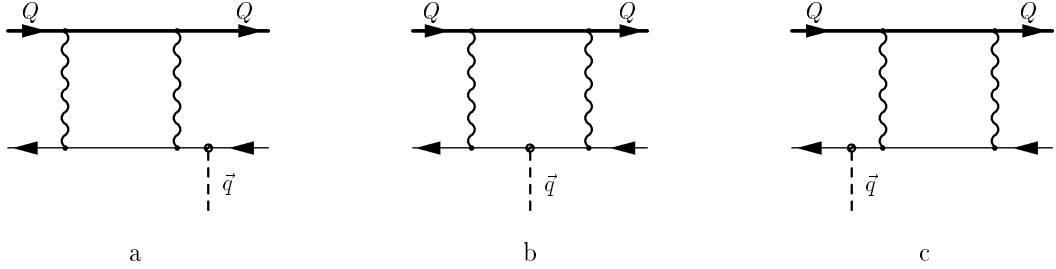,width=14cm}}
 \end{center}
 \caption{
Order-$\alpha_s^2$ diagrams determining the asymptotics of the light
quark formfactor of a heavy hadron. Similar diagrams with the twisted
gluon lines are not shown.} \label{fig3} \end{figure}

A closer look reveals, however, that the cancellation of the leading
$1/|\vec{q}\,|^3$ asymptotics does not hold already at the one-loop
level. The asymptotics has actually the form
$\sim \as^2(\vec{q}\,)/|\vec{q}\,|^3$ which emerges from
the diagrams shown
in Figs.~3 (other diagrams decrease faster in the Feynman gauge). This
leads to a nonzero anomalous dimension of the color-straight operators
at order $\as^2$ and their mixing with color-octet operators. In
particular, the evaluation of the one-loop amplitudes leads to
\beq
{\cal A}_\Gamma (\vec{q}\,) =
 \frac{\frac{\pi^2}{2} \as^2}{|\vec{q}\,|^3}
\left[\left(1-\frac{1}{N_c^2}\right)
\frac{\matel{\tilde H_b}{\bar{b}b\,\bar{q}\Gamma q}{H_b}}{2M_{H_b}}
\:+\;
N_c\left(1-\frac{4}{N_c^2}\right)
\frac{\matel{\tilde H_b}{\bar{b} t^a b\,\bar{q}\Gamma t^a
q}{H_b}}{2M_{H_b}} \:\right]\,.
\label{asym}
\eeq
Eq.~(\ref{44}) then yields for the UV part of the four-fermion operator
\beq
O_{\mbox{\tiny s-s}}\;=\; \frac{\as^2}{4}
\left[\left(1-\frac{1}{N_c^2}\right)\bar{b} b\,\bar{q}\Gamma q\:+\;
N_c\left(1-\frac{4}{N_c^2}\right)
\bar{b} t^a b\,\bar{q}\Gamma t^a q\:\right]\cdot
\ln{\frac{\Lambda_{\rm uv}}{\mu}} \; + \;\mbox{finite piece ,}
\label{uvlog}
\eeq
and likewise for the spin-triplet operators.

A direct calculation of the two-loop anomalous dimensions confirms this.
The computational details are described in Appendix~2. Here we only
quote the result. Let us denote~\footnote{The anomalous dimensions of
the operators are often defined with the opposite sign. We prefer to
use this convention where the meaning of the anomalous and canonical
dimensions are the same. That is, the scaling properties of the
operators are given by the sum (rather than difference) of
their canonical and anomalous dimensions.}
\bea
\mu\frac{\rm d}{{\rm d}\mu}
\left( \begin{array}{c}
O^i \\ T^i \end{array}\right) =
\hat\gamma \left( \begin{array}{c} O^i \\ T^i \end{array}\right) \;.
\label{two66}
\eea
Then
{\small
\bea
\gamma_{11}\! &=&  4 \pi^2  \left(1-\frac{1}{N_c^2}\right)
\left(\frac{\as}{4\pi}\right)^2 +{\cal O}(\as^3)\,, \quad\qquad\qquad
\gamma_{12} \! = \! 4\pi^2 N_c \left(1-\frac{4}{N_c^2}\right) \!
\left(\frac{\as}{4\pi}\right)^2 \!\!+\!{\cal O}(\as^3)
\nonumber\\
\gamma_{21} \!\! &=& \!\! \!\pi^2 N_c \left(1-\frac{1}{N_c^2}\right)
\!\left(1-\frac{4}{N_c^2}\right)\!
\left(\frac{\as}{4\pi}\right)^2 \!\!+\!{\cal O}(\as^3) , \quad
\gamma_{22} = \; 3 N_c \frac{\as}{4\pi} +{\cal O}(\as^2) \:.\qquad
\label{2loop}
\eea }
We note that since $\gamma_{11}$,
$\gamma_{12}$ and $\gamma_{21}$ vanish to order $\as$, these two-loop
anomalous dimensions do not depend on the renormalization scheme. For
$\gamma_{22}$ the second-order terms depend on the scheme and we do not
consider them.

Additional terms are present for the flavor-singlet operators: for
the vector current only $\gamma_{22}$ is modified,
$\gamma_{22} \to \gamma_{22} -\frac{4}{3}
n_f\frac{\as}{4\pi}$. If the operator has the flavor-singlet axial
current then only the diagonal anomalous dimension for the
color-straight operator changes, $\gamma_{11} \to \gamma_{11}
-6 n_f\left(N_c-\frac{1}{N_c}\right)\left(\frac{\as}{4\pi}\right)^2$.

It is interesting that, although $\gamma_{11}$, $\gamma_{12}$ and
$\gamma_{21}$ already appear in the second loop, they are universal. In
particular, they are the same for both timelike and spacelike
components of the light quark currents. A priori this does not need to
hold. We expect that this universality will be violated in the next
order in $\alpha_s$.

The two-loop anomalous dimensions are enhanced, they
contain a large factor $\pi^2$.  Neglecting them introduces
a numerical uncertainty in the running of operators.
We can estimate it by simply setting $\ln\frac{\mu'}{\mu}$ to unity.
The corresponding corrections at $\as=1$ constitute about
$15$~to~$30\%$.  This provides additional justification for the
standard choice of $\as(\mu)=1$ as the low (hadronic) normalization
scale.

We point out that the naive estimate of the power of the asymptotics
$1/\vec{q}^{\,4}$ of the light current formfactors existing in the
literature \cite{luty} is not correct: the actual fall off is only
$1/|\vec{q}\,|^3$ as shown in Eq.~(\ref{asym}),
which, however, is generated only by the exchange of
two gluons with momenta $\sim \vec{q}$. (The modification for the
spin-triplet operators is obvious). This asymptotics can be easily RG
improved using relations Eq.~(\ref{44}), (\ref{46}). To the NLO it
amounts to adding the factor
$\left[\as(\vec{q}\,)/\as(\mu)\right]^{-3N_c/2\beta_0}$  (or
$\left[\as(\vec{q}\,)/\as(\mu)\right]^{(-3N_c+4/3n_f)/2\beta_0}$ for the
flavor-singlet vector current) in front of the color-octet operator,
with $\mu$ the normalization point of the operators.
Since the gap
between the typical hadronic mass $\mu_{\rm had}$ and $m_b$ is not too
large in the logarithmic scale, this observation has, probably, rather
theoretical than practical significance.

It turns out that in the relativistic system the Coulomb interaction is
still strong enough to make the wavefunction density not approaching a
literal constant at zero separation but having the small logarithmic
dependence on distance which appears only at the level of loop
corrections, at order $\sim \as^2$. \vspace*{.25cm}

Based on the application to the lifetimes of heavy hadrons (first of
all, in $B$ mesons) and routine application of factorization, a
standard choice for the basis for four-fermion operators ascending to
the original papers on the subject was \cite{WA,Ds}
\beq
O_{\rm singl} \;=\; \left(\bar{b} \Gamma^{(1)} q \right) \left(\bar{q}
\Gamma^{(2)} b \right) \;, \qquad \qquad O_{\rm oct} \;=\;
\left(\bar{b} t^a \Gamma^{(1)} q \right) \left(\bar{q} t^a
\Gamma^{(2)} b \right)
\label{55}
\eeq
($t^a=\frac{\lambda^a}{2}$, $\lambda^a$ are the usual Gell-Mann color
matrices), that is the $s$-channel color-singlet and color-octet
operators. It appears, however, that a better choice is to classify the
operators according to the color structure in the $t$-channel:
\beq
O \;=\;  \left(\bar{b}_i \Gamma^{(1)} q^j \right)
\left(\bar{q}_k \Gamma^{(2)} b^l \right) \cdot \delta_{il} \delta_{kj}
\;, \qquad \qquad T \;=\; \left(\bar{b}_i \Gamma^{(1)} q^j
\right) \left(\bar{q}_k \Gamma^{(2)} b^l \right) \cdot t^a_{il}
t^a_{kj} \;.
\label{57}
\eeq
In the large-$N_c$ limit these two
bases coincide (up to permutation):
\bea
& &O_{\rm singl} =  2 T +
\frac{1}{N_c} O\;, \qquad \qquad \quad O_{\rm oct} =
\frac{1}{2}\left(1-\frac{1}{N_c^2}\right) O - \frac{1}{N_c} T
\qquad \\
& &O =
2O_{\rm oct} + \frac{1}{N_c} O_{\rm singl}\;, \qquad\qquad T =
\frac{1}{2}\left(1-\frac{1}{N_c^2}\right) O_{\rm singl } -
\frac{1}{N_c}  O_{\rm oct} \;.\qquad
\label{59}
\eea
The $t$-channel octet operators $T$ also
diagonalize the one-loop anomalous dimension matrix; its value depends
on the type of the current, flavor-singlet or octet \cite{vslog}.

We parametrize these generic expectation values encountered in actual
weak decays as
\bea \frac{1}{2M_B}\matel{B}{O_V}{B} &=\; \omega_V \qquad\;
\frac{1}{2M_B}\matel{B}{T_V}{B} &=\; \tau_V \\
\frac{1}{2M_B}\matel{B}{O_A}{B} &=\; \omega_A \qquad\;
\frac{1}{2M_B}\matel{B}{T_A}{B} &=\; \tau_A\;.
\label{wt}
\eea
The parameters $\omega$, $\tau$ have dimension $m^3$ and are constants
in the heavy quark limit. They can be valence or non-valence; the
flavor of the light quark in the operator will be indicated as a
superscript.

For $\Lambda_b$-baryons we denote
\beq
\frac{1}{2M_{\Lambda_b}}\matel{\Lambda_b}{O_V}{\Lambda_b}\;=\;
\lambda \qquad\;
\frac{1}{2M_{\Lambda_b}}\matel{\Lambda_b}{T_V}{\Lambda_b}\;=\;
-\frac{2}{3}\lambda' \;;
\label{baryon}
\eeq
in the valence approximation $\lambda'=\lambda$. The values of
$\lambda$ for $u$ and $d$ quarks are equal, likewise for
$\lambda'$. These valence expectation values will be normally used
without flavor index.

A remark is appropriate to conclude the discussion of the perturbative
renormalization. Strictly speaking, the flavor-singlet operators
can be renormalized in somewhat different ways depending on the
prescription to treat the tadpole-type closed loops.
The free quark loop by dimensional counting scales with the
UV cutoff $\mu$ like $\mu^3$, and describes a possible power mixing
with the $D=3$ ``unit'' heavy-quark operator $\bar{b}b$ already at
order $\as^0$. Although for practically relevant operators such a
``bare'' mixing vanishes for the usual way to regulate the light quark
loop, one can raise the question where this freedom is
reflected in relations (\ref{44},\ref{46}) for a generic $\Gamma$. The
resolution is rather straightforward: the flavor-singlet current
$\bar{q} \Gamma q$ also requires regularization of the closed fermion
loop and, {\em a priori} admits mixing with the unit operator (the
tadpole graph).  This operator does not lead to any physical transition
at $q\ne 0$ but to the forward amplitude with
$\vec{q}=0$. A formally defined current $\bar{q} \Gamma q$ may thus
lead to an additional term proportional to $\delta^3(\vec{q}\,)$ in the
transition amplitude ${\cal A}(\vec{q}\,)$, which would reproduce the
tadpole term in the expectation value.

Similarly, strictly speaking one could have chosen an arbitrary
convention for the phases of the states $\state{\tilde{H}_b(\vec{q}\,)}$
with different momenta $\vec{q}\,$. This would redefine the phase of
the transition amplitude ${\cal A}(\vec{q}\,)$. In our relations such a
freedom was eliminated by adopting Eqs.~(\ref{26},\ref{27},\ref{30})
which ensures, for example, the proper analytic properties of the
transition amplitudes.

In the purely perturbative calculations one can, in principle, consider
not only the actual physical amplitudes, but also similar transition
amplitude induced by the light quark currents carrying color. Applying
to them relations similar to Eq.~(\ref{44}) and (\ref{46}) one would
need to consider the color-nonsinglet quark {\it in} or {\it out}
states. This case requires certain care since such amplitudes may have
additional (gauge-dependent) infrared singularities.

It is worth reiterating that in our analysis it is assumed that all
heavy quark operators are renormalized at a scale well below $m_b$,
which implies a nontrivial -- even if finite -- renormalization
when passing from the full QCD fields. In particular, the vector
$\bar{b}b$ and axial $\bar{b} \vec{\sigma} b$ currents both do not
renormalize in this domain; however, they run differently when evolved
down from the scale $\sim m_b$. While $\bar{b}\gamma_0 b \;\ra
\;\bar{b}b$ is not renormalized, the short-distance renormalization of
$\bar{b}\gamma_k\gamma_5 b \;\ra\; \bar{b} \sigma_k b$ slightly
suppresses it:
$$
\zeta\; \simeq \; 1-\frac{2\as}{3\pi} \,+\, {\cal O}(\as^2)
$$
(the second-order correction has been also calculated \cite{czarm}).
This is not, however, the only short-distance contribution differentiating
the renormalization which, in general, depends on the exact form of the
operators. We will not further dwell on these corrections in our
numerical analysis.

\subsection{Estimates of the color-straight expectation values in $B$
mesons}

Relations (\ref{44},\ref{46}) open a possibility for an alternative
evaluation of the expectation values of the color-straight operators.
It requires knowledge of the light-quark-current formfactors of heavy
hadrons. The direct experimental information about them is scarce.
Therefore, we have to assume a reasonable model. Our general strategy
for all expectation values of interest is the same: decompose the
transition amplitude into the invariant formfactors, and adopt a model
for the formfactors satisfying known constraints.

For the family of $\Lambda_b$ baryons, the number of possible
amplitudes is limited due to the fact that the light degrees of
freedom are spinless -- one can construct only scalar, vector and
tensor currents while pseudoscalar and axial amplitudes vanish.
There are no axial analogues of the expectation values in
Eq.~(\ref{baryon}).
For mesons all amplitudes are possible.
Our main attention will be
devoted to the vector and axial currents, due to the chiral invariance
of phenomenologically relevant four-fermion operators. We do not
consider the tensor current, and only briefly comment on the scalar
one.

There is only one formfactor for the vector current (for each flavor
content) for both $B$ and $\Lambda_b$ describing the only
nonvanishing timelike component:
$$
\matel{B(\vec{q}\,)}{J_\mu}{B(0)}\;=\; -v_\mu F_B(q^2)
$$
\beq
\matel{\Lambda_b(\vec{q}\,)}{J_\mu}{\Lambda_b(0)}\;=\; v_\mu
F_{\Lambda_b}(q^2) \bar u(v,s') u(v,s)\;.
\label{63}
\eeq
One important constraint on the formfactors is their value at $q^2=0$.
The values of $F_{B, \Lambda_b} (0)$ are fixed by the corresponding
charge of the hadron: it is $1$ for the current of a valence quark, and
zero for a `sea' light flavor.\footnote{We adopt the convention where
$B$ mesons have the quark content $b\bar{q}$.}
For the amplitude in Eqs.~(\ref{63}) the
integration over $\vec{q}$ yields
\beq
v_\mu\int\; \frac{{\rm d}^3\vec{q}}{(2\pi)^3}\: F(q^2)\; = \;
\frac{v_\mu}{4\pi^2}\:\int_0^{\infty} \, dt\: \sqrt{t} F(-t)\;.
\label{64}
\eeq

The valence formfactors are expected to decrease for non-zero
$\vec{q}^{\,2}$. For the isovector formfactor the slope at $q^2=0$
(related to the corresponding charge radius) can be estimated in
terms of experimentally observable quantities by an analogue of the
Cabibbo-Radicati sum rule for heavy hadrons \cite{cp}:
\beq
\frac{{\rm d} F(q^2)}{{\rm d} q^2}_{\vert_{q^2=0}}\; = \;
\frac{1}{8\alpha Q^2}\, \sum_{\rm exc} (2J+1)
\frac{\Gamma(B_{\rm exc}\ra B\gamma)} {|\vec k\,|^3}
\label{65}
\eeq
where the sum runs over excitations of $B$ with spin $J$ and $Q$ is the
light quark charge in units of $e$ (a similar relation holds for
baryons as well), and the nonresonant contributions are neglected. This
slope is not yet known experimentally well enough.

Relations following from Eq.~(\ref{64}) can be used in the lattice
simulations to evaluate the expectation values, by measuring the
transition formfactors in a few kinematic points and interpolating
between them. This type of lattice measurements can be simpler than
for the heavy-quark current transitions, since the heavy quarks remain
at rest and the momenta involved in the process do not scale
with $m_Q$. This makes the static approximation rather
straightforward.

If we represent the formfactor as a sum over singularities in the
$t$-channel
\beq
F(q^2)\;=\; \sum_n\: \frac{c_n M_n^2}{M_n^2-q^2} \;,
\label{67}
\eeq
the integral Eq.~(\ref{64}) takes the form
\beq
\int\; \frac{{\rm d}^3\vec{q}}{(2\pi)^3}\: F(q^2)\; = \;
-\frac{1}{4\pi}\: \sum_n \, c_n M_n^3\;\,;
\label{68}
\eeq
we have an additional constraint
\beq
\sum_n \, c_n M_n^2 \;=\; 0
\label{69}
\eeq
following from the fact that the transition amplitudes decrease faster
than $1/\vec{q}^{\,2}$.

It is natural to consider the simplest model of saturation containing
only two lowest-lying $1^-$ states with appropriate isospin quantum
numbers. For example, for $I=1$ we use $\rho(770)$ and $\rho(1450)$.
With fixed normalization at $q^2=0$ and the constraint (\ref{69}) this
model predicts the value of the integral in terms of the two masses. It
is worth noting that such a model would obviously lead to equal
expectation values of the operators in $B$ and $\Lambda_b$. Imposing an
additional constraint from the slope of the formfactors would allow one
to fix all residues in a three-pole model as well, which can be hoped
to yield a more accurate estimate.

A word of reservation is in order at this point. Such a saturation of
the nucleon formfactors by two lowest $t$-channel resonances is known
to provide a good approximation for moderate $q^2$ where the
experimental formfactors are described by the double-pole expressions.
There is no general theoretical justification for such a coincidence,
and more resonances are expected to play a role for larger $q^2$. In
particular, at $-q^2 \gsim 1 \GeV^2$ the formfactor can decrease
faster. Due to the phase space factor the role of the domain of large
$q^2$ is enhanced.  The contribution of higher states, while affecting
a little the formfactors near $q^2=0$, still can significantly change
the integral (\ref{68}). We will return to this point later.

It is clear that since the asymptotics of the amplitudes has an odd
power of $1/|\vec{q}\,|$, their representation by a finite number of
the $t$-channel resonances is not possible. The true spectrum of the
$t$-channel states must extend to arbitrary high masses. It applies
even if there were no typical for QCD $\log$-like dependence of the
asymptotics $1/|\vec{q}\,|^3$. It does not affect our estimates since
we evaluate the operators in the effective theory where the
high-momentum component of the hadrons is peeled off.

Addressing the color-straight operators containing the axial
light-quark current (which does not vanish in $B$ mesons) we note that
its matrix elements
are generally described by two formfactors, just as in the well-known
case of spin-$\frac{1}{2}$ fermions. These are analogues of the
axial-charge and weak magnetism terms. Spontaneous breaking of the
chiral symmetry modifies the value of the axial-charge formfactor from its
symmetric limit of 1 at $q^2=0$. Nevertheless, for the
isovector current, its conservation $\partial_\mu J_{\mu 5}(x)=0$ in the
chiral limit leads to a relation between the formfactors, so that
only one, the axial-charge formfactor is independent, as in the case of
the vector current. At $q^2=0$ this relation equates the axial-charge
formfactor to the $B^*B\pi$ coupling $g$  (the heavy-quark analogue of
the Goldberger-Treiman relation). Given the value of $g$, therefore, one
can evaluate the expectation value exactly as outlined for the
vector currents.

For the isosinglet axial current, one has to take into consideration
the anomalous term, the topological charge density $Q$:
\beq
\partial_\mu J_{\mu 5}^{(0)}(x)\;=\; 2i\bar{q}\hat m_q\gamma_5 q(x)
\,+\,n_f Q(x)\;, \qquad Q(x)\:=\;\frac{\as}{4\pi}\,
{\rm Tr} \,G_{\alpha\beta}\tilde{G}^{\alpha\beta}(x)
\label{72}
\eeq
with $\hat m_q$ the light quark mass matrix.
The matrix elements of $Q$ over the $B$ meson states are not known, and
the above relation appears to be less constraining. In the large-$N_c$
limit the difference between singlet and nonsinglet formfactors is
expected to disappear; however, the practical validity of this
approximation for the anomalous term is questionable. These problems
are addressed in the next section.

Let us briefly mention the case of the scalar current. Although the
corresponding formfactor is not fixed at $q^2=0$, its value for the
valence quarks can be obtained from the $SU(3)$ mass splittings:
\beq
\frac{1}{2M_B}\matel{B^+}{\bar{u} u(0)}{B^+} \simeq
\frac{M_{B_s}-M_B}{m_s} \simeq  0.7 \,,
\quad
\frac{1}{2M_{\Lambda_b}}\matel{\Lambda_b}{\bar{u} u(0)}{\Lambda_b} \simeq
\frac{M_{\Xi_b}-M_{\Lambda_b}}{m_s} \simeq 1.4 \,,
\label{74}
\eeq
This estimate is obtained with the help of the Zweig rule
in a similar way as done in \cite{sigma} to extract the value of the
nucleon $\sigma$-term from the $SU(3)$ splittings in the baryon octet.
We neglected here the light quark masses $m_{u,d}$ and took
$m_s(1\GeV)\simeq 130 \MeV$. The mass of the baryon $\Xi_b$ has not
been yet measured; the above estimate used the prediction
$M_{\Xi_b}=5805.7\pm 8.1$ MeV \cite{xib}.
The normalization point dependence of $m_s$ in these relations
reproduces the dependence of the scalar current.

In what follows we will apply the described strategy to evaluation of a
few expectation values of operators with the vector and axial light
quark currents. First, however, we discuss qualitatively the saturation
of the integral of the formfactors in the adopted models.

In the case of the (valence) vector current we have $F(0)=1$. It is
natural to think also that $|F(q^2)| < 1$ at spacelike $q$. Let us
further assume that $F(q^2)$ is small enough above a certain scale
$\mu$, so that we can neglect it there:
\beq
F(q^2) \simeq 0 \qquad \mbox{ at } \qquad -q^2 > \mu^2\;.
\label{76}
\eeq
Then we get an upper bound
\beq
\left|
\frac{1}{2M_B} \matel{H_b}{\bar{b}\gamma_\mu b \,
\bar{q} \gamma_\mu q}{H_b}
\right| \; < \; \frac{\mu^3}{6\pi^2} = 0.017\mbox{ GeV}^3
\label{78}
\eeq
for $\mu=1$ GeV.
This bound is of the type discussed in \cite{boost}; the numerical
coefficient coincides with the one given there.

Are assumptions like Eq.~(\ref{76}) reasonable? In the effective theory
with the normalization point $\mu$ the momenta of fields exceeding
$\mu$ are absent, whether or not the full theory yields a logarithmic
`tail' at large momenta. For example, it is not possible to exchange a
gluon with momentum $|\vec{q}\,| >\mu$ in such a theory. The exact shape
of the formfactor would depend on the concrete realization of the
effective theory. The amplitude may not vanish exactly due to multiple
gluon exchanges with $|\vec{q}\,| < \mu$, however would then decrease
exponentially.

On the other hand, a step-like formfactor saturating the bound
(\ref{78}) is clearly unrealistic. Therefore, we can assume instead
that
\beq
|F(q^2)| \;<\; {\rm e}\, ^{-\vec{q}^{\,2}/\mu^2}\;,
\label{79}
\eeq
which results in
\beq
\left|
\frac{1}{2M_B} \matel{H_b}{\bar{b}\gamma_\mu b \,
\bar{q} \gamma_\mu q}{H_b}
\right| \; < \; \frac{\mu^3}{8\pi^{3/2}} = 0.022\mbox{ GeV}^3
\label{80}
\eeq
with the same value for $\mu$ as in (\ref{78}).
As a matter of fact, an exponential ansatz
${\rm e}\, ^{-\vec{q}^{\,2}/\mu^2}$ for the formfactor with $\mu^2$
adjusted to reproduce the `charge' radius, is a reasonable model for the
possible behavior of the
valence formfactor of purely soft degrees of freedom.
In particular, if we adopt the slope at $q^2=0$ following from the
pole model
keeping only the two lowest
states with masses $M_1$ and $M_2$ (this is a good approximation in
known cases) and use the exponential formfactor, the resulting
integrals appear noticeably smaller than in the two-pole ansatz
itself:
\beq
\frac{1}{2M_B} \matel{H_b}{\bar{b}\gamma_\mu b \,\bar{q} \gamma_\mu
q}{H_b}  \; = \; F(0) \frac{1}{8\pi^{3/2}}
\left(\frac{M_1^2 M_2^2}{M_1^2+M_2^2}\right)^{3/2}\;,
\label{81}
\eeq
$$
\mu^2= \frac{M_1^2 M_2^2}{M_1^2+M_2^2}\;.
$$

The above discussed bounds rely on the assumption that $|F(q^2)| <1\,$.
It always holds in nonrelativistic QM, however we do not know a
general rigorous proof in QCD. For the isovector current one can employ
the equal-time commutation relation ($J_\mu^a$ can be both the
vector $V_\mu^a =\bar q\gamma_\mu \frac12\tau^a q$ or the axial current
$A_\mu^a =\bar q\gamma_\mu\gamma_5 \frac12\tau^a q$)
\beq
\left[J_0^+(\vec x\,),J_0^-(\vec y\,)\right]\;=\; \delta^3(\vec x-\vec y)\,
2V_0^3(\vec x\,)
\label{82}
\eeq
to represent $1-|F(q^2)|^2$ at $q^2 < 0$ as a difference of two sums of
the distinct transition probabilities:
\beq
|F(\vec{q}^{\,2})|^2\;=\; 1\,-\,
\left( \sum_n\, |F_n^+|^2\:-\:\sum_m\, |F_m^-|^2 \right)\;.
\label{83}
\eeq
Here $|F_n^+|^2$, $|F_m^-|^2$ schematically denote the transition
probabilities in, say, $B^-$ meson induced by the currents $\bar{u}
\gamma_0 d$ and $\bar{d} \gamma_0 u$ with the momentum transfer
$\vec{q}$, respectively (and similarly for $\Lambda_b$). In the second
sum only the states with $I=\frac{3}{2}$ contribute. Since there are no
valence $\bar d$ quarks in $B^-$, in the large-$N_c$ limit the last term
with the wrong sign would vanish. Also, in this limit the isoscalar meson
formfactor is expected to coincide with the isovector one. Therefore,
the large-$N_c$ arguments allow to establish such a QM bound for
all formfactors of interest.

There is no natural normalization for the axial (pseudoscalar)
formfactors at small momentum. Moreover, the amplitudes generally have
an enhancement due to the pion pole. However, the domain $\vec{q}^{\,2}
\sim m_\pi^2$ yields a very small contribution to the integral (see,
e.g., Eq.~(\ref{78})). The significant contribution can
originate only from momenta $\gsim 1 \GeV$ where one expects the
effects of chiral symmetry breaking to become insignificant.
As noted above, the equal-time commutation relation (\ref{82})
can be used to derive a sum rule of the type (\ref{83}) also for the
matrix elements of the axial isovector current. Its explicit form is similar
to (\ref{83}) and reads
\beq
|G_1(\vec{q}^{\,2})|^2\;=\; 1\,-\,
\left( \sum_n\, |G_n^+|^2\:-\:\sum_m\, |G_m^-|^2 \right)
\label{83axial}
\eeq
with $G_1(q^2)$ defined below in (\ref{112}) and $|G_n^+|^2$, $|G_m^-|^2$
are the analogues of the $F_n$ amplitudes for transitions induced by
the axial current acting on a $B$ meson.  At $q^2=0$ this sum rule is
just the familiar Adler-Weisberger sum rule and the amplitudes $G_n$ are
related to pion couplings between the ground and excited states. The
explicit form of these sum rules for heavy mesons and baryons can be
found in \cite{cp,AW}.  Therefore, we expect the type of bounds
(\ref{78},\ref{80}) to hold also for the axial current expectation
values as well.

\section{Numerical estimates}

In this section we estimate the expectation values of the
color-straight four-fermion operators relevant for the lifetimes of
beauty hadrons. The light quark fields are left-handed; the Penguin
diagrams bring in the right-handed fields as well. Nevertheless, the
chiral structure of the currents admits only the vector or axial
light quark currents. Since the coefficient functions can include
the momentum of the decaying $b$ hadron (its velocity), the timelike and
spacelike components  enter, in general, with different weights; the
three-dimensional rotation invariance is still preserved. Finally,
since the forward matrix elements are considered, only
the parity-conserving (three-dimensional) scalar expectation values
survive. Therefore, we need to consider the operators
\beq
O_V=(\bar{b}b)(\bar{q}\gamma_0q)= (\bar{b}\gamma_\mu
b)(\bar{q}\gamma_\mu q)\;, \qquad
O_A =-(\bar{b}\vec\sigma b)(\bar{q}\vec\gamma \gamma_5 q)=
(\bar{b}\gamma_\mu \gamma_5 b)(\bar{q}\gamma_\mu \gamma_5q)\;.
\label{84}
\eeq
As was mentioned, the expectation value of the operator $O_A$ in
$\Lambda_b$ vanishes.

\subsection{Vector current}

We first consider the case of $B$ mesons.
Assuming only isospin symmetry, we define the isovector and
isoscalar four-quark matrix elements by
$$
\frac{1}{2M_B}\matel{B_i}{\bar{b}\gamma_\mu b \,
\bar{q} \tau^a\gamma_\mu q}{B_j} \;=\; V_3 \,\tau^a_{ij}
$$
\beq
\frac{1}{2M_B}\matel{B_i}{\bar{b}\gamma_\mu b\,
\sum_{q=u,d} \bar{q}\gamma_\mu q}{B_j} \;=\; V_1 \,\delta_{ij}
\label{85}
\eeq
The indices $i,j$ label the respective state in the isospin doublet
$i=(\bar d,-\bar u)$.
Accordingly, we introduce the isospin-triplet and singlet vector
formfactors
$$
\frac{1}{2M_B}\matel{B_i(\vec{q})}{\bar{q} \tau^a\gamma_\mu
q}{B_j(0)} \;=\; v_\mu \,{\cal F}_3(q^2)\, \tau^a_{ij} $$ \beq
\frac{1}{2M_B}\matel{B_i(\vec{q})}{\sum_{q=u,d} \bar{q}\gamma_\mu q}
{B_j(0)} \;=\; -v_\mu\, {\cal F}_1(q^2) \, \delta_{ij}\;,
\label{86}
\eeq
with the normalization conditions ${\cal F}_1(0)={\cal F}_3(0)=1\,$.
Using the two-pole ansatz saturated by $\rho(770)$ and $\rho(1450)$ for
the nonsinglet current ${\cal F}_3$ , we get from Eqs.~(\ref{67}-\ref{69})
\beq
V_3 \;\simeq\; \frac{1}{4\pi} \frac{M_1^2 M_2^2}{M_1+M_2} \;\simeq \;
0.045 \GeV^3\;.
\label{87}
\eeq
It is natural to saturate the $I=0$ formfactor ${\cal F}_1$ by the states
$\omega(782)$ and $\omega(1420)$. It then leads to almost the same
numerical estimate for $V_1$
as for $V_3$. The reason is obviously an almost exact
degeneracy of the vector states in the isovector and isosinglet
channels. Although it perfectly fits the large-$N_c$ picture, we cannot
be sure what is the actual accuracy of such a conclusion. Nevertheless,
in view of such a suppression of the difference, we will take the two
different combinations of the expectation values which actually
parametrize the valence and non-valence contributions:
\beq
\frac{V_1-V_3}{2}\;=\;
\frac{1}{2M_B}\matel{B^-}{\bar{b}\gamma_\mu b \,
\bar{u}\gamma_\mu u}{B^-} \;\simeq\; -0.044\GeV^3
\label{88}
\eeq
$$
\frac{V_1+V_3}{2}\;=\;
\frac{1}{2M_B}\matel{B^-}{\bar{b}\gamma_\mu b \,
\bar{d}\gamma_\mu d}{B^-} \;\simeq\; {\cal O}\left(10^{-4}
\GeV^3\right) \;.
$$
The last number, clearly, is at best an order of magnitude estimate.

We can try to estimate the violation of the $SU(3)$ flavor symmetry
considering the expectation value of $\bar{b}\gamma_\mu b \,
\bar{s}\gamma_\mu s$ in $B_s$ mesons. For this we
 saturate the formfactor with the
vector $\bar{s}s$ states $\phi(1020)$ and $\phi(1680)$,
which corresponds to the ``ideal'' mixing in the $\omega-\phi$ system
\cite{PDG}.  In this case we would get
\beq
\frac{1}{2M_B}\matel{B_s}{\bar{b}\gamma_\mu b \,
\bar{s}\gamma_\mu s}{B_s} \;\simeq\; -0.085\GeV^3\;,
\label{89}
\eeq
i.e. almost twice larger than the first estimate (\ref{88}).

A closer look reveals, however, that the above expectation values are
saturated at rather high momenta. Half of the `valence' value comes
from $|\vec{q}\,| > 1.5 \GeV$, and from even higher momenta in $B_s$.
For this reason these estimates exceed the bounds (\ref{78}),
(\ref{80}) discussed in the previous section,
for a reasonable scale $\mu \simeq 1 \GeV$. For example,
adopting the exponential ansatz for the formfactor, we would get
$$
\frac{1}{2M_B}\matel{B^-}{\bar{b}\gamma_\mu b \,
\bar{u}\gamma_\mu u}{B^-} \;\simeq\; -0.007\GeV^3
$$
\beq
\frac{1}{2M_B}\matel{B_s}{\bar{b}\gamma_\mu b \,
\bar{s}\gamma_\mu s}{B_s} \;\simeq\; -0.015\GeV^3 \;.
\label{90}
\eeq

A somewhat unexpected result of these simple estimates is the
apparently large amount of $SU(3)$ breaking in Eqs.~(\ref{88},\ref{89}).
While it is not clear to what extent this is an artifact of our use of the
simple two-pole ansatz for the formfactors over a wide domain of $q^2$,
it is worth noting that a simple mechanism exists which could account
for it. It is well-known that the isovector charge radius of a hadron
diverges in the chiral limit \cite{bz}. This indicates that the contribution of
the low-momentum region in the integral over the formfactor (\ref{64})
is more suppressed in nonstrange B mesons compared to the
B$_s$ case. Since the two-pole model does not capture the origin
of this phenomenon (the contribution of the
two-body $\pi\pi$ intermediate state in the $t$-channel),
it is conceivable that the magnitude of $SU(3)$ violation in
the matrix elements does exceed a few percent.

It is interesting to compare the above estimates with the evaluation
based on vacuum factorization. Both types of estimates have the same
${\cal O}\left(N_c^0\right)$  scaling in $N_c$. However, the vacuum
contribution for the color-straight operators is ${\cal
O}\left(N_c^0\right)$, similar to other meson states. This is in
contrast to the case of the $t$-channel octet operators where the
vacuum state is $N_c$-enhanced. Therefore, the factorization estimate
is not expected to give an accurate result. For the valence expectation
value one has
\beq
\frac{1}{2M_B}\matel{B^-}{\bar{b}\gamma_\mu b \,
\bar{u}\gamma_\mu u}{B^-}_{\rm factor} \;=\; - \frac{1}{4N_c}
\tilde f_B^2(\mu)M_B
\label{91}
\eeq
(the non-valence value vanishes). Here $\tilde f_B$ denotes the
annihilation constant of $B$ for the $\bar{b} \gamma_\alpha\gamma_5 u$
current normalized at a low point $\mu$ where factorization must be
applied \cite{fb,vslog}, in contrast to the physical $f_B$ defined
for the current normalized at $\mu \gg m_b$:
\beq
\tilde f_B(\mu) \;\simeq \; f_B
\left[\frac{\as(\mu)}{\as(m_b)}\right]^{-\frac{2}{\beta_0}}\;.
\label{92}
\eeq

The physical value of $f_B$ lies, probably, around $160\MeV$.
However, to the leading order in $1/m_b$ we work in, it is more
consistent to use the asymptotic value which differs from the
physical one by  $1/m_b$ and non-logarithmic perturbative corrections.
These decrease the physical value of $f_B$ by about $20\%$ \cite{braun}.
Therefore, we adopt $\tilde f_B=160\MeV$ for $\as(\mu)=1$, yielding
\beq
\frac{1}{2M_B}\matel{B^-}{\bar{b}\gamma_\mu b \,
\bar{u}\gamma_\mu u}{B^-}_{\rm factor} \;\simeq\;
- 0.011 \GeV^3\;,
\label{93}
\eeq
which is significantly lower than Eq.~(\ref{88}).

The fact that the corrections to factorization can be significant, is
expected. Unfortunately, there are good reasons to question the
accuracy of the alternative estimate (\ref{88}) either, and a too large
$SU(3)$ breaking is another indication. We think that it is justified
to consider the estimate (\ref{88}) for the valence expectation value
rather as an upper bound, while the number obtained in the
exponential ansatz a reasonable lower bound.  A conservative
estimate then is
$$ \frac{1}{2M_B}\matel{B^-}{\bar{b}\gamma_\mu b \,
\bar{u}\gamma_\mu u}{B^-} \;=\; -(0.025\pm 0.015)\GeV^3
$$
\beq
\frac{1}{2M_B}\matel{B^-}{\bar{b}\gamma_\mu b \,
\bar{d}\gamma_\mu d}{B^-} \;\approx \; {\cal O}\left(5\cdot
10^{-4}\GeV^3 \right) \;.
\label{94}
\eeq
Similar estimates can be adopted for strange quarks in $B_s$.
\vspace*{0.25cm}

Next we turn to baryons. Under the light flavor $SU(3)$ group the
$\Lambda_b$ and $\Xi_b$ states transform as an antitriplet
$T_i=(\Xi_b^d, -\Xi_b^u, \Lambda_b)$.
In the limit of $SU(3)$ symmetry there are only two independent
formfactors, which can be defined as
$$
\frac{1}{2M_{\Lambda_b}}\matel{T_i(\vec{q}\,)}{\bar{q}
\lambda^a\gamma_\mu
q}{T_j(0)} \;=\; v_\mu \,{\cal F}^\Lambda_8(q^2)\, \lambda^a_{ji}
\,\bar u(v,s') u(v,s)
$$
\beq
\frac{1}{2M_{\Lambda_b}}\matel{T_i(\vec{q}\,)}{\sum_{q=u,d,s}
\bar{q}\gamma_\mu q} {T_j(0)} \;=\; v_\mu \, {\cal F}^\Lambda_1(q^2)
\, \delta_{ij} \, \bar u(v,s') u(v,s)\;.
\label{97}
\eeq
The normalization at $q^2=0$ is ${\cal F}^\Lambda_1(0)=2$,
${\cal F}^\Lambda_8(0)=-1$. Using a similar model for the formfactors
as in the meson case we get the same expectation values
(up to the sign)  for the valence matrix elements, and strongly
suppressed non-valence contributions. For example,
the two-pole model yields the following value for the $\Lambda_b$
matrix elements
\beq
\frac{1}{2M_{\Lambda_b}}\matel{\Lambda_b}{\bar{b}\gamma_\mu b \,
\bar{u}\gamma_\mu u}{\Lambda_b}=
\frac{1}{2M_{\Lambda_b}}\matel{\Lambda_b}{\bar{b}\gamma_\mu b \,
\bar{d}\gamma_\mu d}{\Lambda_b}
\simeq
\left\{
\begin{array}{ll} 0.007\GeV^3 & \mbox{\small (exponential)} \\
0.045\GeV^3 & \mbox{\small (two-pole)}
\end{array} \right.
\label{99}
\eeq
For the same reasons as before it is natural to consider the two-pole
value as an upper bound. The expectation values of the non-strange
operators in the $\Xi_b$ states emerge the same as in (\ref{99}),
whereas
$\matel{\Xi_b}{\bar{b}\gamma_\mu b \, \bar{s}\gamma_\mu s}{\Xi_b}$
again literally appears twice larger than in the $SU(3)$ limit. As
discussed in the $B$ meson case, such a large symmetry violation can be
suspected to be, at least partially, an artifact of the two-pole model.

It is worth noting that in the case of heavy baryons the light quark
formfactors have anomalous Landau thresholds associated with the
$N\bar{N}B$ triangle diagrams. It is well known that it is such
singularities that determine the low-momentum behavior of the
formfactors and, in particular, the large charge radius of weakly-bound
states like deuteron \cite{anomal}. For the $\Lambda_b$ formfactor the
anomalous singularity starts at
\beq
t_{\rm thr} = 4M_N^2\left(1 - \frac{(M_{\Lambda_b}^2-M_N^2-M_B^2)^2}
{4M_B^2 M_N^2}\right) = 3.2 \GeV^2\;.
\label{102}
\eeq
In this system, however, the corresponding mass still lies higher than
the states we use to saturate the formfactors. Moreover, there is no
reason to expect the residues to be significant (for example, they are
$1/N_c$ suppressed). Therefore, we believe that these singularities do
not play a role in the expectation values we study. In any case, a
refined estimate will be possible with a better knowledge of the
formfactors, say using determination of its slope based on the
application of the Cabibbo-Radicati sum rule to the radiative decays of
excited baryons.

\subsection{Operators with axial current}

The expectation value of the operators $\bar{b} \gamma_\mu\gamma_5b\,
\bar{q} \gamma_\mu\gamma_5 q$ vanishes in the $\Lambda_b$ baryon
family, and we consider it only for $B$ mesons employing the relation
Eq.~(\ref{46}).  In this case
\beq
(\vec{S}_b \vec{\epsilon}\,)\,  \state{B(\vec{q}\,)}\;=\;
\state{B^*(\vec{q}, \vec{\epsilon}\,)}\;, \qquad
\vec{S}_b\;=\; \int\, {\rm d}^3x\: \bar{b} \vec\gamma\gamma_5 b(x)\;.
\label{110}
\eeq
Since the light degrees of freedom carry spin $\frac{1}{2}$, the axial
current is parametrized by two formfactors; the third possible
structure has wrong $T$ parity and vanishes. This is an exact analogue
of the absence of the second-class currents in $\beta$-decays of light
baryons. Thus one has
$$
\frac{1}{2M_B}\matel{B^*_i(\vec{q}, \vec{\epsilon}\,)}
{\sum_{q=u,d,s} \bar{q}\gamma_\mu \gamma_5 q(0)}{B_j(0)} \;=\;
\left\{ \epsilon^*_\mu G_1^{(0)}(q^2)- (\epsilon^* q) q_\mu G_0^{(0)}(q^2)
\right\} \delta_{ij}\;,
$$
\beq
\frac{1}{2M_B}\matel{B^*_i(\vec{q}, \vec{\epsilon}\,)}{
\bar{q} \lambda^a\gamma_\mu \gamma_5 q(0)}{B_j(0)} \;=\;
\left\{ \epsilon^*_\mu G_1(q^2)- (\epsilon^* q) q_\mu G_0(q^2)
\right\}  \lambda^a_{ji}
\;.
\label{112}
\eeq
Absence of the structure $(\epsilon^* q) v_\mu$ is easy to show
explicitly (note that in any case the timelike component of the axial
current does not enter the four-fermion operators). Using
Eq.~(\ref{110}) and the fact that $\vec{S}_b$ commutes with all
light-quark field operators, we get an equality
$$ \matel{B^*(\vec{q}, \vec{\epsilon}\,)}
{J_{\mu 5}(0)}{B(0)}^* = \matel{B(\vec{0})}
{J_{\mu 5}(0)}{B^*(\vec q, \vec{\epsilon}\,)} =
\matel{B^*(\vec{0}, \vec{\epsilon^*})}{J_{\mu 5}(0)}{B(\vec q\,)}\,.
$$
Inserting here the formfactor decompositions (\ref{112}) for these
matrix elements
and taking into account the fact that $G_i$ are real from $T$
invariance, one finds that the structure $(\epsilon q) v_\mu$ appears
with opposite signs on the two sides of the equality. Hence its
coefficient must vanish.

We thus get
\beq
\frac{1}{2M_B}\matel{B^-}{\bar{b}\gamma_\mu \gamma_5 b
\,(\bar{u}\gamma_\mu \gamma_5 u - \bar{d}\gamma_\mu \gamma_5
d)}{B^-} \;=\;
-\frac{1}{4\pi^2}\, \int_0^{\infty}\, {\rm d}t\, \sqrt{t}
\left(3G_1(-t)+ t G_0(-t) \right)
\label{113}
\eeq
and a similar expression for the singlet matrix elements.

In the chiral limit, which will be assumed in what follows, the
isovector formfactor $G_1$ at $q^2=0$ is related to the $BB^*\pi$
coupling:
\beq
G_1(0)\;=\;g\;.
\label{114}
\eeq
The nonrelativistic quark model predicts $g=-0.75$. However, the QCD
sum rules estimates yield lower values \cite{qcdg1,qcdg2,qcdg3,qcdg4}.
The recent analyses predict $g=-0.3$ \cite{qcdg2,qcdg3}
which is consistent with the
existing experimental bounds $g^2 = 0.09-0.5$ \cite{gexp}.  Moreover,
the equation of motion $\partial_\mu J_{\mu 5}=0$ leads to
$q^2G_0(q^2)= G_1(q^2)$ at all $q^2$, therefore for the nonsinglet
expectation value Eq.~(\ref{113}) takes the form
\beq
\frac{1}{2M_B}\matel{B^-}{\bar{b}\gamma_\mu \gamma_5 b
\,(\bar{u}\gamma_\mu \gamma_5 u - \bar{d}\gamma_\mu \gamma_5
d)}{B^-} \;=\;
-\frac{1}{2\pi^2}\, \int_0^{\infty}\, {\rm d}t\, \sqrt{t}
G_1(-t) \;.
\label{115}
\eeq

The only nonvanishing contribution to $G_0$ from the pseudoscalar
states in the isovector channel comes from the
massless pion. The $J^{PC}=1^{++}$ states contribute to both $G_1$ and
$G_0$:
$$
G_1(t) \;=\; \sum_n\, \frac{g_n M_n^2}{M_n^2-t}
$$
\beq
G_0(t) \;=\; -\frac{g}{m_\pi^2-t}\: +\: \sum_n\,
\frac{g_n}{M_n^2-t}
\label{116}
\eeq
with the condition $\sum_n g_n=g$ replacing the zero-transfer
normalization of the vector formfactor. A faster than $1/q^2$ fall-off
of the transition amplitude requires additionally
\beq
\sum_n\, g_n M_n^2\;=\;0\;,
\label{118}
\eeq
which is analogous to the second constraint in Eq.~(\ref{69}) for the
vector current.

In the numerical estimates for the isotriplet current we will consider
both a two-pole ansatz for $G_1(q^2)$
and the exponential ansatz
\beq
G_1(q^2)= g{\rm e}\,^{-\vec{q}\,^2/\mu^2}\;,
\qquad G_0(q^2) = g\frac{{\rm e}\,^{-\vec{q}^{\,2}/\mu^2}}{q^2} \;
\qquad \mbox{ with } \qquad
\mu^2=\frac{M_1^2 M_2^2}{M_1^2+M_2^2}\;.
\label{119}
\eeq
Such a choice of $\mu$ ensures that the two ans\"atze have the same
behavior at small $q^2$.

In the $I,J^{PC}=1,1^{++}$ channel only the lowest-lying state $a_1(1260)$
has been observed. For the numerical estimates we will need also
the mass of its first radial excitation $a'_1$.
This has been extracted in \cite{DoGo} from an analysis of the
Weinberg sum rules.
The value obtained in \cite{DoGo} for the mass of the $a'_1$ resonance
is 1869 MeV, which is what we will use in our estimates.

First, with the two-pole ansatz we obtain
\beq
\frac{1}{2M_B}\matel{B^-}{\bar{b}\gamma_\mu \gamma_5 b
\,(\bar{u}\gamma_\mu \gamma_5 u - \bar{d}\gamma_\mu \gamma_5
d)}{B^-} \;=\;
-\frac{g}{2\pi}\,\frac{M_1^2 M_2^2}{M_1+M_2} \;\simeq\;
0.084 \GeV^3 \;.
\label{121}
\eeq
As explained above, we have adopted in this estimate the value $g=-0.3\,$
\cite{qcdg2,qcdg3}. For the exponential formfactor one obtains
a smaller value
\beq
\frac{1}{2M_B}\matel{B^-}{\bar{b}\gamma_\mu \gamma_5 b
\,(\bar{u}\gamma_\mu \gamma_5 u - \bar{d}\gamma_\mu \gamma_5
d)}{B^-} \;=\;
-\frac{g}{4\pi^{3/2}}\,\left(\frac{M_1^2
M_2^2}{M_1^2+M_2^2}\right)^{3/2} \;\simeq\; 0.015  \GeV^3 \;.
\label{122}
\eeq

In a completely analogous way one can estimate the matrix element
of the $I=0$ octet axial current with the flavor content of $\eta$.
With the mass of the state $f_1(1285)$ close to mass of $a_1$ and
assuming a similar degeneracy for the second excitation we do not get
appreciable $SU(3)$ violation and, therefore, obtain for
$$
\frac{1}{2M_B}\matel{B^-}{(\bar{b}\gamma_\mu \gamma_5 b)
(\bar{u}\gamma_\mu \gamma_5 u + \bar{d}\gamma_\mu \gamma_5 d
- 2\bar{s}\gamma_\mu \gamma_5 s)}{B^-}
$$
the same value as in Eqs.~(\ref{122}) and (\ref{121}).

In the case of the singlet axial current we need to account for the
presence of the anomalous term in its divergence,
$$
\partial_\mu J_{\mu 5}^{(0)}(x)\;=\; n_f \, Q(x)\:, \qquad
Q(x)\;=\; \frac{\as}{4\pi} G\tilde{G}(x)\;.
$$
For simplicity, we will assume the exact $SU(3)$ chiral limit.
The value of the isosinglet axial formfactor at small $q^2$ is given
by the matrix element of the anomalous divergence $Q(x)$:
\beq
\frac{1}{2M_B}\matel{B^*(\epsilon, \vec{q}\,)}{Q(0)}
{B(0)} \;=\; \frac{i}{n_f}G_1^{(0)}(0) (\epsilon^* q) \;+\; {\cal O}(q^2)\;.
\label{131}
\eeq
We used here the fact that there are no massless particles in the
singlet channel and  consequently $G_0^{(0)}$ is finite.
The contribution of the pseudoscalar states to the formfactor
$G_0^{(0)}$ does not vanish and is determined by their coupling to
$Q(x)$, $\matel{n}{Q(0)}{0}$. Similarly, the $G_1^{(0)}$ and
$G_0^{(0)}$ formfactors at arbitrary $q^2$ are not directly related to
each other, but the difference $i(G_1^{(0)}-q^2G_0^{(0)})$ equals to
the matrix element of $n_f Q(x)$.  Very little is known directly about
these flavor-singlet expectation values or $B^*B\eta^{(\prime)}$
coupling.

Nevertheless, for estimates one usually employs an approximation
in which
the matrix elements of $Q(x)$ are saturated by the $\eta'$ pole.
Moreover, the couplings of the whole nonet of the pseudoscalar mesons
$\pi,\, K,\, \eta,\, \eta'$ are assumed $SU(3)$-symmetric.
This assumption is incorporated in the simple $\sigma$-models proved to
be successful in describing the properties of light hadrons. This model
\cite{Vene,DyEi}
naively has an $U(3)\times U(3)$ chiral symmetry; the $U(1)$ problem is
solved by adding the anomalous term with $Q(x)$ and assuming the
nonvanishing (in the quenched approximation, that is, in QCD without
light flavors) value of the zero-momentum correlator of the topological
charge densities $Q(x)$
\beq
\lambda^4\;=\; \int \, {\rm d}^4 x \:
\matel{0}{iT\{Q(x)\,Q(0)\}}{0}_{n_f=0}
\label{140}
\eeq
which leads, basically, to the nonzero anomalous mass of $\eta'$ meson
$m_{\eta'}^2 = \lambda^4/f_\pi^2$. Adopting such a model, we also have
$G_1^{(0)}(0) \simeq G_1(0)$.

A possible justification for such a picture lies in the large-$N_c$
approximation. However, in this limit $m_{\eta'}^2 \propto 1/N_c$ and
the anomalous $U(1)$ symmetry effectively restores, which
seems not be close to actual world where the anomalous mass of $\eta'$ is
numerically large and the octet-singlet mixing in pseudoscalars is
small.  It is probable that there exists a deeper dynamic reason
explaining the practical validity of such approximation.

The model with a single $\eta'$ state in the pseudoscalar channel which
merely shifts the pole in the nonsinglet amplitudes from $q^2=0$ to
$m_{\eta'}^2$, while describing reasonably well the low-$q^2$ matrix
elements of the topological charge density, leads to their too mild
suppression at large $q^2$. In reality they are expected to decrease
very fast above a typical momentum scale of the nonperturbative vacuum
configurations. In order to mimic this behavior, we have to employ at
least two pseudoscalar states saturating the correlators of $Q$, and we
take the state $\eta'$ with a mass of $M_{\eta'}=1295\MeV$ as the
second pole.  One expects an $J^{PC}=0^{-+}$ $SU(3)$ singlet in this
mass region, accompanying the observed octet of pseudoscalars
containing $\pi(1300),\eta(1295)$.  In reality, the wide `gluonium'
states can give a significant contribution.  Probably, an exponential
ansatz is a better approximation here.

In principle, the spectrum of the axial-vector singlet states has no
direct relation to the anomaly and the $U(1)$ problem. Hence we take
for the corresponding masses the experimental values, namely
$f_1(1285)$ and its first radial excitation $f'_1$, neglecting
their mixing with the octet states.
$f_1(1285)$ lies close to the isotriplet state $a_1(1260)$, indicating
smallness of the annihilation effects.
Therefore we will take for the mass of the first radial
excitation $M_{f'_1}=M_{a'_1}\simeq 1870\MeV$ in the numerical
estimates below.
In the two-pole model we have
$$
G_1^{(0)}(q^2) \;=\; G_1^{(0)}(0) \frac{M_{f_1}^2 M_{f'_1}^2}
{(M_{f_1}^2-q^2)(M_{f'_1}^2-q^2)}
$$
\beq
G_1^{(0)}(q^2)- q^2 G_0^{(0)}(q^2) \;=\;
G_1^{(0)}(0) \frac{M_{\eta'}^2 M_{\eta'(1295)}^2}
{(M_{\eta'}^2-q^2)(M_{\eta'(1295)}^2-q^2)}\,.
\label{141}
\eeq
The last equation replaces the second of Eqs.~(\ref{116}).

As a result, the difference in the estimates compared to the isotriplet
current lies basically in the anomalous term, and is not too
significant. Numerically, we get for the two-pole ansatz
$$
\frac{1}{2M_B}\matel{B^-}{\bar{b}\gamma_\mu \gamma_5 b \,
\sum_{q=u,d,s}\bar{q}\gamma_\mu \gamma_5 q}{B^-}
\;=\;
-\frac{G_1^{(0)}(0)}{2\pi}\,\left(\frac{M_{f_1}^2 M_{f'_1}^2}
{M_{f_1}+M_{f'_1}}
+\frac{1}{2}
\frac{M_{\eta'}^2 M_{\eta'(1295)}^2}{M_{\eta'}+M_{\eta'(1295)}}
\right)
$$
\beq
\;\simeq\;-(0.29 + 0.054) G_1^{(0)}(0) \GeV^3= 0.1 \GeV^3
\qquad  \mbox{ at }
\; G_1^{(0)}(0)=-0.3\;.
\label{142}
\eeq
In the numerical estimate we used the equality
$G_1^{(0)}(0)=G_1(0)=g$ which holds in the large-$N_c$
limit, as discussed above.

We present also a calculation of the matrix element (\ref{142})
employing the exponential ansatz. This is constructed in the same way as
for the axial charge formfactor. For the topological charge
formfactor we use an exponential normalized at $q^2=0$ by the same
value $G_1^{(0)}(0)$ and vary the slope parameter
$\mu_Q^2$ from $\mu_Q^2=m_{\eta'}^2=0.92 \GeV^2$
(corresponding to the pole dominance by $\eta'$ alone)
to $0.59\GeV^2$ (corresponding to the two-pole model,
see Eq.~(\ref{81})).
This yields the following numerical estimate:
$$
\frac{1}{2M_B}\matel{B^-}{\bar{b}\gamma_\mu \gamma_5 b
\,\sum_{q=u,d,s}\bar{q}\gamma_\mu \gamma_5 q}{B^-} \;\simeq\;
-\frac{G_1^{(0)}(0)}{4\pi^{3/2}}\,\left( \left[\frac{M_{f_1}^2
M_{f'_1}^2}{M_{f_1}^2+M_{f'_1}^2}\right]^{\frac{3}{2}}
+\frac{1}{2} \mu_Q^3
\right)
$$
\beq
\;\simeq\;-(0.053 + (0.010 \mbox{ to }  0.020)) G_1^{(0)}(0) \GeV^3
\simeq
(0.02 \mbox{ to } 0.023) \GeV^3
\label{144}
\eeq
with the same value for
$G_1^{(0)}(0)$ as before.  One could try to estimate the effects of
$SU(3)$ breaking by accounting for the known shifts in masses and
mixing. We think, however, that such models are too crude to capture
correctly details of $SU(3)$ violation, and we do not attempt it here.

Combining the above results
for the octet and singlet expectation values,
we get the following estimates for the valence and non-valence
axial expectation values:
\begin{eqnarray}
\label{147}
\frac{1}{2M_B}\matel{B^-}{\bar{b}\gamma_\mu \gamma_5 b
\,\bar{u}\gamma_\mu \gamma_5 u}{B^-} &\simeq&
0.018 \GeV^3\\
& & 0.09 \GeV^3\nonumber\\
\label{148}
\frac{1}{2M_B}\matel{B^-}{\bar{b}\gamma_\mu \gamma_5 b
\,\bar{d}\gamma_\mu \gamma_5 d}{B^-} &\simeq&
0.002 \GeV^3\\
& & 0.007 \GeV^3\nonumber
\end{eqnarray}
where the upper (lower) value corresponds to the exponential (two-pole)
formfactor model.

We note that the effect of the axial anomaly can be numerically important,
although it is formally of order $1/N_c$. In the approximations considered
here, it makes a contribution of about $15\%$ of the total singlet
expectation values (\ref{142}) and (\ref{144}), respectively, and it
can dominate the non-valence matrix elements.

Finally, we quote also the factorization approximation estimate for the
valence expectation value of the axial current. We obtain
\beq
\frac{1}{2M_B}\matel{B^-}{\bar{b}\gamma_\mu \gamma_5 b \,
\bar{u}\gamma_\mu \gamma_5 u}{B^-}_{\rm factor} \;=\;
\frac{3}{4N_c} \tilde f_B^2(\mu)M_B
\;\simeq\; 0.034 \GeV^3\;\;.
\label{150}
\eeq
Just as for the color-straight operators containing the vector current,
we do not expect the factorization approximation to be accurate.
However, it is interesting that the factorization value for the axial
operators is less suppressed compared to the case of the vector
current, and appears to be closer to the estimates given above.

\subsection{Estimates from the fourth sum rule}

One of the color-octet operators, the flavor-singlet vector four-fermion
operator can be estimated in an alternative way. This operator
$\tilde{O}_D$
\beq
\tilde O_D\; =\; \sum_{q=u,d,s}(\bar b\gamma_\mu t^a b) (\bar
q\gamma_\mu t^a q)
\eeq
is the QCD generalization
of the Darwin term in atomic physics \cite{motion,optical,Darwin}:
\beq
\frac{1}{2M_{H_b}} \matel{H_b}{2\pi\as \tilde O_D }{H_b}\;=\;
-\frac{1}{2M_{H_b}} \matel{H_b}{O_D }{H_b}\;=\;
- \left(\rho_D^3\right)_{H_b}\;.
\label{dar}
\eeq
On the other hand, it determines the third moment of the small velocity
(SV) structure function, the so-called fourth sum rule, and is related
to quantities measurable in the semileptonic decays. For example, in $B$
mesons this sum rule in the resonant approximation takes the form
\cite{Darwin}
\beq
\frac{1}{3} \rho_D^3 \;=\;
E_{1/2}^3 |\tau_{1/2}|^2 + 2E_{3/2}^3 |\tau_{3/2}|^2 +
\cdots\;,
\label{4thsr}
\eeq
where $\tau_j$ are so-called ``oscillator strengths'' which determine
the small velocity transition amplitudes into the excited $p$-wave
states with spin of light degrees of freedom $j$.
The excitation energies of these states with respect to
the ground state $s$-wave mesons are denoted with $E_j$ (for a recent
discussion see review \cite{rev}).

It should be noted that the literal application of the fourth sum rule
requires specific regularization scheme for the operators. In view of
the tentative nature of our estimates we neglect these subtleties here.
Some of them were discussed in \cite{optical} and more recently in
\cite{dipole} and \cite{varenna}. We only mention that the
large negative logarithmic anomalous dimension of the Darwin operator
\cite{vsold}
\beq
(\alpha_s \tilde O_D)_{\mu'} \; \simeq \;
\left(\frac{\alpha_s(\mu')}{\alpha_s(\mu)} \right)^{13/(2\beta_0)}
(\alpha_s \tilde O_D)_{\mu}\;, \qquad \;\;
\beta_0=\frac{11}{3}N_c-\frac{2}{3} n_f
\eeq
to a large extent offsets the apparent scale dependence in evaluation of
$\aver{\tilde O_D}$ via $\rho_D^3$ due to the $\mu$-dependence of
$\as$ (or, similarly, the scale dependence of the factorization
estimate of $\rho_D^3$ in $B$ mesons). According to the standard
practice we use for our estimates the scale corresponding to
$\as(\mu)=1$.

The recent discussion of the status of the sum rule evaluation of
$\rho_D^3$ in $B$ mesons can be found in the review \cite{rev},
Sect.~4. The corresponding value is in a reasonable agreement with the
factorization estimates (see Appendix~3, Eq.~(\ref{a37})).

Since the straightforward factorization cannot be used for baryons, we
will apply the fourth sum rule to evaluate the operator $\tilde{O}_D$ in
$\Lambda_b$. The fourth sum rule for it takes the form
\beq
\frac{1}{2M_{\Lambda_b}} \matel{\Lambda_b}
{2\pi\alpha_s \tilde O_D}{\Lambda_b}
= -3 \sum_n E_n^3 |\sigma^{(n)}|^2\,.
\label{4bar}
\eeq
The states appearing in the r.h.s.\ are orbital excitations of the
$\Lambda_b$ baryon with quantum numbers of the light degrees of freedom
$s_\ell^{\pi_\ell}=1^-$. Their excitation energies are $E_n$ and
$\sigma^{(n)}$ are the corresponding oscillator strengths describing
semileptonic decays of $\Lambda_b$ to the analogous excitations of
the $\Lambda_c$ baryon; they are defined as in \cite{IWY}.

The first excited states appearing in this sum rule have been identified
as the doublet of negative-parity baryons
$\Lambda_{c_1}^+(2593)$ and $\Lambda_{c_1}^+(2625)$.
Unfortunately no experimental information is available to date on the
transition amplitude $\sigma^{(1)}$ governing the decays of
$\Lambda_b$ into both of them, although it will be ultimately measured.

The important piece of information would be the slope
$\rho^2_{\Lambda_b}$ of the elastic IW function in $\Lambda_b$.
This quantity is more accessible than $\sigma^{(1)}$ and will be measured
in the near future at LEP. In the absence of the data we can use the
second sum rule (Voloshin's ``optical'' sum rule) \cite{volopt} for
$\bar\Lambda_{\Lambda_b}=M_{\Lambda_b}-m_b \simeq M_{\Lambda_c}-m_c$.
As discussed in \cite{rev} (for earlier application see also
\cite{Darwin,third}), we can estimate $\rho_D^3$ using just the
excitation energy of the first states. We simply take
$M_{\Lambda_c^*}\simeq 2.615\GeV$ and the first excitation energy
$\Delta_1\simeq 330 \MeV$. Assuming $\bar\Lambda_B\simeq 600\MeV$ and,
therefore, $\bar\Lambda_{\Lambda_b}\simeq 900\MeV$, we then have
\beq
\left(\rho_D^3\right)_{\Lambda_b}\;\approx \;
\frac{3}{2}\, \Delta_1^2 \bar\Lambda_{\Lambda_b}\;\approx\; 0.15
\GeV^3\;.
\label{darwin}
\eeq
(A similar estimate $\left(\mu_\pi^2\right)_{\Lambda_b} \simeq
\frac{3}{2}\, \Delta_1 \bar\Lambda_{\Lambda_b} \simeq 0.45 \GeV^2$
agrees well with the mass relations \cite{buvprep,Ds} for charm and
beauty in the meson and baryon sectors). We note in passing that, most
probably, the large mass of the heavy baryon implying larger
$\bar\Lambda$ compared to $B$ meson is due to larger slope
$\rho^2_{\Lambda_b}$; the higher-dimension operators, therefore, can be
even smaller than in mesons.

The sum of the expectation values for all three light flavors
$2\lambda'+\lambda'_s$ is related to the expectation value of
the Darwin operator \cite{motion}:
$$ 2\lambda'+\lambda'_s\;=\; \frac{3}{4\pi\as}
\left(\rho_D^3\right)_{\Lambda_b}\;.
$$
Hence, we estimate the $SU(3)$-singlet color-octet expectation value in
$\Lambda_b$ as
\beq \lambda'_{u+d+s}\;\approx \; \frac{3}{4\pi}
\left(\rho_D^3\right)_{\Lambda_b}\;\approx \; 0.036\GeV^3
\label{octlam}
\eeq
with uncertainty about $30$--$40\%$. The estimate of $\lambda'_u $ can
be obtained if we neglect the small contribution of the non-valence
strange quarks:
\beq
\lambda'_{u}\;=\; \lambda'_{d}\;\simeq\;
\frac{3}{8\pi} \left(\rho_D^3 \right)_{\Lambda_b} \;\simeq \;
0.018\GeV^3 \;.
\label{lam'}
\eeq
We note that we get a reasonable agreement with the
quark model relation $\lambda' \approx \lambda$ between the straight
and octet expectation values in $\Lambda_b$ for our central estimates,
Eq.~(\ref{99}).
It is interesting that the
corresponding value of the diquark density at origin appears
close to our central estimate for mesons $-\omega_V$ (but larger than
the alternative analogue of $|\Psi(0)|^2$ in mesons $\tilde f_B^2M_B/12$).
It also exceeds the estimates obtained in the QCD sum
rules \cite{rQCD} or quoted from bag models \cite{HYC}; these analyses
determined the combination $\frac{4}{3} \lambda'-\frac{1}{3} \lambda$
which generally emerged in the ball park of $0.004\GeV^3$.

\section{Nonfactorizable pieces in the matrix elements of the
four-quark operators}

As was mentioned earlier, there are four operators (for a given light
quark flavor) determining the corrections to the mesons widths. These
are color-straight $O$ and color-octet $T$. Each of these can contain
either timelike (vector $O_V$, $T_V$) or spacelike (axial
$O_A$, $T_A$) components of light and heavy quark currents.

The color-octet expectation values $\tau$ in general can be estimated
using vacuum factorization, since the operators $T$ coincide with the
operators colorless in the $s$-channel up to $1/N_c$ terms
(see Eq.~(\ref{59})). For such operators vacuum factorization is expected
to work up to $1/N_c$ corrections. This gives
\bea
\tau_V &=& -\frac{\tilde f_B^2(\mu) M_B}{8}
\left(1-\frac{1}{N_c^2}\right) \simeq -0.015 \GeV^3\\
\tau_A &=& \frac{3\tilde f_B^2(\mu) M_B}{8}
\left(1-\frac{1}{N_c^2}\right) \simeq 0.045 \GeV^3\,.
\eea

It is interesting to note that the leading $1/N_c$ corrections to the
factorization approximation can be
estimated in a phenomenological approach. For this the expectation
value of the color singlet $O_{\rm singl}$ in (\ref{59}) is written as
\bea
\langle H_b|(\bar b\Gamma q)(\bar q\Gamma b)|H_b\rangle =
\langle H_b|\bar b\Gamma q|0\rangle \langle 0|\bar q\Gamma b|H_b\rangle
+\sum_n \langle H_b|\bar b\Gamma q|n\rangle \langle n|\bar q\Gamma b|H_b\rangle
\,.
\eea
The leading corrections to the vacuum factorization approximation are
of order 1 and come from one-particle intermediate states like
$\pi\,(\eta,\eta')$, $\rho\,(\omega)$,
$a_1$ for $B$ mesons, or light baryons for $\Lambda_b$.
The corresponding transition amplitudes have been evaluated in the QCD sum
rules \cite{bbraun} and lattice simulations with an accuracy sufficient for
determining the scale of the effects.
Alternatively, they can be approximately obtained from the
corresponding decays of charmed particles.

Large-$N_c$ arguments
{\it per se} do not
ensure that the vacuum factorization approximation works in the
case of the color-straight operators, for the nonfactorizable
contribution appears at the same order in $N_c$ as the factorizable one.
Their Wilson coefficients are not suppressed compared to those of the
color-octet operators $T$ (see Table~1), and they can be
important even if their expectation values are formally subleading in
$1/N_c$. Moreover, the factorizable part of the expectation values has
only a specific Lorentz structure which is subject to the strong chirality
suppression $\sim m_c^2/m_b^2$ in the effects of weak annihilation (WA)
in mesons.  Nonfactorizable contributions there can be dominant
\cite{WA,Ds}.

Nonfactorizable effects also appear as the expectation values of the
non-constituent quark operators. Although they do not split the widths
of $B^{\pm}$ and $B^0$, they can differentiate the meson vs. baryon
lifetimes. Numerically they seem to be strongly suppressed, with a
possible exception of the Darwin operator which will be discussed
below.

The nonfactorizable effects in $B$ mesons were first discussed in the
framework of the $1/m$ expansion in \cite{WA} where the
parametrization
\bea\label{gv1}
&
\frac{1}{2M_B}
\matel{B}{\bar b\gamma_\mu (1-\gamma_5)q
\bar q\gamma_\nu (1-\gamma_5) b}{B}\; &=\;
\frac{\tilde{f}_B^2 M_B}{2} (v_{s}v_\mu v_\nu - g_{s} g_{\mu\nu})\\
&
\frac{1}{2M_B}
\matel{B}{\bar b\gamma_\mu (1-\gamma_5)t^a q
\bar q\gamma_\nu (1-\gamma_5)t^a b}{B}\; &=\;
\frac{\tilde{f}_B^2 M_B}{2} (v_{o}v_\mu v_\nu - g_{o} g_{\mu\nu})
\label{gv}
\eea
was suggested motivated by the analysis of the WA effects: neglecting
the $c$ quark mass WA is governed by $g_o$ and $g_s$. (In the
factorization approximation $v_s=1$ and $v_o=g_o=g_s=0$.) These
parameters are related to $\omega$, $\tau$ in the following way:
\bea
\tilde{f}_B^2 M_B\, v_s\;&=&\; -\frac{2}{N_c} \omega_V
\,+\,\frac{2}{3N_c} \omega_A \,-\, 4\tau_V\,+\, \frac{4}{3} \tau_A\\
\tilde{f}_B^2 M_B\, g_s\;&=&\; -\frac{1}{N_c} \omega_V
\,-\,\frac{1}{3N_c} \omega_A \,-\, 2\tau_V\,-\, \frac{2}{3} \tau_A  \\
\tilde{f}_B^2 M_B\, v_o\;&=&\;
- \left(1-\frac{1}{N_c^2}\right) \omega_V
\,+\, \frac{1}{3} \left(1-\frac{1}{N_c^2}\right) \omega_A
\,+\, \frac{2}{N_c} \tau_V\,-\, \frac{2}{3N_c} \tau_A \\
\tilde{f}_B^2 M_B\, g_o\;&=&\;
-\frac{1}{2} \left(1-\frac{1}{N_c^2}\right) \omega_V
\,-\, \frac{1}{6} \left(1-\frac{1}{N_c^2}\right) \omega_A
\,+\, \frac{1}{N_c} \tau_V\,+\, \frac{1}{3N_c} \tau_A\,.\qquad
\label{wt2gv}
\eea
The inverse relations expressing $\omega$ and $\tau$ via $v$ and $g$
are given in Appendix~3.

The color counting rules suggest that $\tau_{V,A}\sim N_c$ while
$\omega_{V,A}\sim N_c^0$. The factorization estimates for $\tau_{V,A}$
in the large-$N_c$ limit are expected to hold with the $1/N_c$
accuracy. Therefore, knowledge of the color-straight operators allows
to estimate the leading, $1/N_c$ terms in $v_o$ and $g_o$:
\bea
 v_o\;&\simeq&\;-
\frac{3\omega_V - \omega_A}{3\tilde{f}_B^2 M_B}
\; -\; \frac{1}{2N_c} \qquad \mbox{(valence)}\\
 g_o\;&\simeq&\;
-\frac{3\omega_V + \omega_A}{6\tilde{f}_B^2 M_B} \;.
\label{gvo}
\eea
(The term $-1/(2N_c)$ is absent for non-valence expectation values.)

Since the non-valence expectation values appear to be suppressed, we
only consider the valence matrix elements generically denoted by the
superscript $(v)$. Let us consider for definiteness the charged $B$
meson; the corresponding parameters are then defined by
Eqs.~(\ref{gv1}),(\ref{gv}) with $q=u$.
Although $\omega_V$ and $\omega_A$ are
not precisely evaluated, we still observe a clear tendency to
cancellations in $g_o$ and, therefore, suppression of the effects of
WA. Say, for the exponential ansatz we get
\beq
v_o^{(v)}\; \approx\; -0.07\;, \qquad\;
g_o^{(v)} \;\approx \; 0.004\;.
\label{expon}
\eeq
For the two-pole ansatz representing the upper limit in our estimates,
we get
\beq
v_o^{(v)}\; \approx\; 0.4\;, \qquad\;
g_o^{(v)} \;\approx \; 0.05\;.
\label{2pole}
\eeq
The nonfactorizable octet parameters seem to emerge
suppressed. In particular, the  expectation value of
the operator responsible for
WA is very small. The color-singlet expectation value $g_s$ was not
estimated in the literature. It is natural to think \cite{Ds} that the
scale of $g_s$ does not exceed that of $g_o$. The above estimates then
illustrate the degree of suppression of the effects of WA when the $c$
quark mass is neglected.

It is appropriate to note at this point that there is a convincing
experimental evidence that WA is indeed strongly suppressed in heavy
mesons. The width  difference between $D_s$ and $D^0$ is very sensitive
to WA. Even though the literal $1/m_c$ expansion in charmed particles
is hardly applicable at the quantitative level, the significance of such
effects would have led to a large $\tau_{D_s}$-- $\tau_{D^0}$
difference. Barring accidental cancellations one gets a typical
estimate \cite{Ds}
$$
|g_o,\,g_s|\; \lsim \; 10^{-2}\;.
$$
We note, therefore, that our estimates, whatever tentative they are,
indicate a strong enough suppression. It is interesting that
the QCD sum rule estimates of the parameters $g$ made in 1992 by
V.~Braun~\footnote{Private communication to N.~Uraltsev in March 1992.
Unpublished.} yielded close values, a few units $\times 10^{-2}$.
Later evaluations gave $v_o\simeq 0.05, \; g_o\simeq 0.05$
\cite{Cherny}, and $v_o\simeq 0.1, \; g_o\simeq 0.03$ \cite{BLLS}; they
were simplified in many aspects, though.
While the expectation value of $v_o$ generally emerges of the order
of $0.05$, our estimates for $g_o$ seem to predict typically
somewhat smaller values $\sim 10^{-2}$, in a better agreement with the
experimental indications.

The non-valence expectation values are probably even further suppressed.
Our estimates yielded $v_o^{(nv)}\approx (0.5\mbox{ to }2)\cdot 10^{-2}$
and $g_o^{(nv)}\approx -(0.25\mbox{ to } 1)\cdot 10^{-2}$, with the
dominant part coming from the axial current via the anomalous terms.

For baryonic expectation values there is no vacuum factorization
approximation. This does not mean, of course, that they are suppressed.
The color-straight
expectation values $\lambda$ were estimated in the preceding sections
to vary from $0.007$ to $0.045\GeV^3$; the interval above $0.03\GeV^3$
seems improbable, though. The estimate of the color-octet
$\lambda'_{u,d}$ based on the evaluation of the Darwin operator yielded
about $0.018\GeV^3$.

A different parametrization of the valence expectation values was used
in \cite{ns2}:
\bea
\label{151}
& &
\frac{1}{2M_B} \matel{B^-}
{(\bar b\gamma_\mu(1-\gamma_5) u)(\bar u\gamma_\mu (1-\gamma_5) b)}
{B} \;= \;\frac{\tilde f_B^2(\mu) M_B}{2} B_1(\mu)\\
& &
\frac{1}{2M_B} \matel{B^-}
{(\bar b(1-\gamma_5) u)(\bar u (1+\gamma_5) b)}{B^-} \;= \;
\frac{\tilde f_B^2(\mu) M_B}{2} B_2(\mu)\\
& &
\frac{1}{2M_B} \matel{B^-}
{(\bar b\gamma_\mu(1-\gamma_5) t^a u)
(\bar u\gamma_\mu (1-\gamma_5) t^a b)}{B^-} \;= \;
\frac{\tilde f_B^2(\mu) M_B}{2} \varepsilon_1(\mu)\\
\label{154}
& &
\frac{1}{2M_B} \matel{B^-}
{(\bar b(1-\gamma_5) t^a u)(\bar u (1+\gamma_5) t^a b)}{B^-} \;= \;
\frac{\tilde f_B^2(\mu) M_B}{2} \varepsilon_2(\mu)\;.
\eea
These parameters are related as follows:
\bea
&
\tilde{f}_B^2 M_B\, B_1\;=\; \tilde{f}_B^2 M_B(v_s-4g_s) \;&=\;
4(\tau_V+\tau_A) \,+\,\frac{2}{N_c} (\omega_V+\omega_A)\\
&
\tilde{f}_B^2 M_B\, B_2\;=\; \tilde{f}_B^2 M_B(v_s-g_s) \;&=\;
-2(\tau_V-\tau_A) \,-\,\frac{1}{N_c} (\omega_V-\omega_A)\\
&
\tilde{f}_B^2 M_B\, \varepsilon_1\;=\; \tilde{f}_B^2 M_B(v_o-4g_o)
\;&=\; -\frac{2}{N_c} (\tau_V+\tau_A) \,+\,
\left(1-\frac{1}{N_c^2}\right) (\omega_V+\omega_A)\\
&
\tilde{f}_B^2 M_B\, \varepsilon_2\;=\; \tilde{f}_B^2 M_B(v_o-g_o)
\; & =\;
\frac{1}{N_c} (\tau_V-\tau_A) \,-\,
\frac{1}{2}\left(1-\frac{1}{N_c^2}\right) (\omega_V-\omega_A)\,.
\qquad \qquad
\eea
The above estimates for the octet expectation values neglecting
$1/N_c^2$ terms look for $\varepsilon_{1,2}$ as
\beq
\varepsilon_1
\;\approx \; -0.085 \mbox{ to } 0.17\;, \qquad\;
\varepsilon_2 \nonumber
\;\approx \; -0.07 \mbox{ to } 0.33\;,
\eeq
while the QCD sum rule calculations
read $\varepsilon_1 \simeq -0.15,\varepsilon_2 \simeq 0$ \cite{Cherny}
and
$\varepsilon_1 \simeq -0.04\pm 0.02,\varepsilon_2 \simeq 0.06\pm 0.03$
\cite{BLLS}

For convenience,
we give in Table 1 the central
estimates of the four-fermion expectation values discussed above in $B$
mesons ($\omega_{V,A}$, $\tau_{V,A}$) and in $\Lambda_b$ ($\lambda$,
$\lambda'$ for a fixed flavor, $u$ or $d$). Since the non-valence
expectation values are strongly suppressed, we do not quote them here.
\begin{table}[t]
\begin{center}
\begin{tabular} {|c|c|c|c|c|c|c|c|c|}
\hline
   & $\omega_V $ & $\omega_A $ & $ \tau_V $ & $\tau_A $  & $ \lambda_u
$ & $\lambda'_u $ & $\left(\rho_D^3\right)_B$ &
$\left(\rho_D^3\right)_{\Lambda_b} $
\\ \hline
Sects. 4.1-4.2 & $-0.020$ & $0.045$ & -- & -- & $0.020$ & -- & -- & -- \\
\hline factorization & $-0.011$ & $0.034$ & $-0.015$ & $0.045$ & -- &
-- & $0.10$ & -- \\ \hline $4^{\rm th}$ sum rule & -- & -- & $-0.028$ &
-- & -- & $0.018$ & $0.18$ & $0.15$ \\ \hline \end{tabular}
\end{center}
\caption{Estimated valence expectation values in $B$ and $\Lambda_b$,
in $\GeV^3$; factorized contributions assume $\tilde f_B=160\MeV$.
\label{TABLE1}}
\end{table}
\vspace*{.2cm}

Concluding this section, we note that
there are two exact positivity constraints on the expectation
values of the $s$-channel colorless operators: $v_s-g_s >1$
($v_s-g_s >0$ for non-valence) and $g_s >0$.

The first inequality follows from inserting a complete
set of intermediate
states $\state{n}$ in the matrix element $\langle B|(\bar b\gamma^0
(1-\gamma_5) q) (\bar q\gamma^0 (1-\gamma_5) b)|B\rangle$. We obtain
\beq
\label{ineq1}
v_s - g_s \;=\; 1 \:+\: \frac{1}{\tilde f_B^2 M_B^2}
\,\sum_n \int \, {\rm d}\mu(n) |\langle n|\bar
q\gamma^0 (1-\gamma_5) b |B\rangle |^2 > 1\,,
\eeq
where ${\rm d}\mu(n)$ stands for the phase space.
The $1/N_c$ contributions in the r.h.s. come when the
intermediate states $n$ are  $\pi$, $\rho$, $a_1$, etc. For
non-valence quarks the vacuum factorization contribution $1$ in the
r.h.s. explicitly showing $\state{n}=\state{0}$
vanishes.\footnote{These inequalities assume a
physical regularization scheme for composite operators in which, for
example, there is a power mixing of the four-fermion operators with the
unit one, $\bar{b}b(0)$. For a recent discussion see, e.g.
\cite{rev,varenna}.}
In terms of the parameters $B_i$, $\varepsilon_i$, the constraint
(\ref{ineq1}) reads $B_2 > 1$.
Estimates of \cite{BLLS} give values for $B_i$ compatible with 1.

The second inequality is obtained by taking spacelike $\mu=\nu=i$ in
(\ref{gv1}).  Summing over $i$ yields
\beq
\label{ineq2}
g_s \;=\;
\frac{1}{3\tilde f_B^2 M_B^2}\sum_n \int \,{\rm d}\mu(n)
\sum_i|\langle n|\bar q\gamma^i (1-\gamma_5) b |B\rangle |^2 \; > \;
0\,.
\eeq
(For the $B$ parameters this is $B_2 > B_1 $.)

A similar inequality can be obtained for the baryonic matrix elements:
\beq
\lambda' - \frac{1}{4}\lambda \;=\;
\frac{3}{4M_{\Lambda_b}}\,
\sum_n\int \, {\rm d}\mu(n)
|\langle n|\bar q(1+\gamma_5)b)|\Lambda_b\rangle|^2 > 0\,,
\label{ineql}
\eeq
where we used the identity
\beq
\langle \Lambda_b|
(\bar b(1-\gamma_5)q)(\bar q(1+\gamma_5)b)|\Lambda_b\rangle =
-\frac{1}{2N_c} \matel{\Lambda_b}{O_V}{\Lambda_b} \,-\,
\matel{\Lambda_b}{T_V}{\Lambda_b} \;.
\eeq
In the constituent quark model the bounds
for $B$ mesons become equalities; the
relation Eq.~(\ref{ineql}) merely expresses
the fact that the diquark wavefunction at origin is positive. It does
not seem to be very restrictive for our estimates.
There is an additional constraint on the expectation values of the
operators with chirality flip for light quarks, however
we are not interested in them.
The above bounds can be refined and even the actual
approximate estimates can be obtained evaluating the contributions of a
few lowest intermediate states
in the hadronic
saturation Eqs.~(\ref{ineq1})-(\ref{ineql}).

\section{Corrections to the decay widths}

In this section we give the expressions for the corrections to the
widths in terms of the effective four-fermion operators normalized at
a low scale. These expressions were originally derived in \cite{vsold}.
We present them here for completeness and book-keeping purposes, in a
more convenient form.

Let us introduce the notation $\Delta
\hat{\Gamma}$ for the operator
describing the corrections to the inclusive width
$\Delta\Gamma_{H_b}$ of the beauty hadron $H_b$:
\beq
\Delta \Gamma\;=\; \frac{1}{2M_{H_b}}
\matel{H_b}{\Delta \hat{\Gamma}}{H_b}\;.
\eeq
In what follows we neglect the effects generated in the KM suppressed
decays which have a factor $|V_{ub}/V_{cb}|^2$, and by Penguin
operators in ${\cal H}_{\rm weak}(\Delta B=1)$ at the scale $m_b$.

Without accounting for the perturbative QCD effects in the domain $k\ll
m_b$ one has {\small
\bea
\nonumber
\Delta \hat{\Gamma}\; &=& \;
\frac{G_F^2m_b^2}{2\pi}\, |V_{cb}|^2 (1-y)^2
\left\{\left(c_1^2+c_2^2+\frac{2}{N_c}c_1 c_2
\right)\left[O_V^u+O_A^u\right]
+4 c_1 c_2 \left[T_V^u+T_A^u\right]\right\}\\
\nonumber
&-& \;
\frac{G_F^2m_b^2}{4\pi}\, |V_{cb}|^2 (1-y)^2
\left\{\left(c_1^2+c_2^2+\frac{2}{N_c}c_1 c_2
\right)\left[(1+y)O_V^{d'}+
\frac{1}{3}(1-y)O_A^{d'}\right]\right.\\
\nonumber
& & \;\; \left.
+\;2(2 c_1 c_2+N_c c_2^2) \left[(1+y)T_V^{d'}+
\frac{1}{3} (1-y) T_A^{d'} \right]\right\}\\
&-& \;
\frac{G_F^2m_b^2}{4\pi}\, |V_{cb}|^2 \sqrt{1-4y}
\left\{\left(c_1^2+c_2^2+\frac{2}{N_c}c_1 c_2\right)\left[O_V^{s'}+
\frac{1}{3}(1-4y)O_A^{s'}\right] \right.\\
\nonumber
& & \;\; \left.
+\; 2(2 c_1 c_2+N_c c_2^2) \left[T_V^{s'}+
\frac{1}{3} (1-4y) T_A^{s'} \right]\right\}\;-\\
\nonumber
& &\hspace{-1.7cm}
\frac{G_F^2m_b^2}{2\pi}\, |V_{cb}|^2
\left\{\left(c_1^2+c_2^2+\frac{2c_1 c_2}{N_c} + \frac{n_\ell}{2N_c}\right)
\left[O_V^{c}+\frac{O_A^{c}}{3}\right]
+2(2 c_1 c_2+N_c c_1^2 + \frac{n_\ell}{2}) \left[T_V^c+
\frac{T_A^c}{3}  \right]\right\}.
\label{bare}
\eea}
We denoted here $y= m_c^2/m_b^2$ and $d'=d\cos\theta_c+s\sin\theta_c$,
$s'=s\cos\theta_c-d\sin\theta_c$.  The
$m_c$-dependence for the operators with external $c$ quark legs
is completely neglected.
(Eventually they will lead only to Penguin-type operators which are
estimated, basically, in the leading-log approximation.)
We included the contribution from the semileptonic decays with
$n_\ell=2$ species of light leptons (the $\tau$ contribution is
suppressed by the phase space).
Since numerically $m_c^2/m_b^2 \approx \mu_{\rm
had}/m_b$, keeping $m_c^2/m_b^2$ corrections apparently is not
legitimate in practice at all. We retain these terms only for getting
an idea of the scale of the finite-$m_c$ corrections in the
coefficient functions.

The perturbative evolution below $m_b$ in the LLA is particularly simple
in this basis: the color-straight operators $O$ do not renormalize. The
color-octet operators $T$ renormalize in a universal way with
$\gamma_T=3N_c$, except for the flavor singlet vector-like operator
similar to $\tilde O_D$ which has anomalous dimension
$\gamma_{\tilde D}= 3 N_c -\frac{4}{3} n_f$ where $n_f$ is the number
of open flavors ($\gamma_D=-\frac{13}{3} N_c$).
At the scale below the charm mass the operators with the
$c$-quark fields merely vanish. As a result, at the low normalization
point $\mu$ we have
$$
\Delta \hat{\Gamma}\; = \;
\frac{G_F^2m_b^2}{2\pi}\, |V_{cb}|^2 \; \times
$$
{\footnotesize
$$
\left\{ (1-y)^2
\left(c_1^2+c_2^2+\frac{2}{N_c}c_1 c_2\right)
\left[O_V^u+O_A^u- \frac{1+y}{2} O_V^{d'} - \frac{1-y}{6}
O_A^{d'} - \frac{\sqrt{1-4y}}{2(1-y)^2} O_V^{s'}
- \frac{(1-4y)^{\frac{3}{2}}}{6(1-y)^2} O_A^{s'}\right]
\right.
$$
$$
\left.  + \; \zeta (1-y)^2 \Big[ 4c_1 c_2
\left(T_V^u+T_A^u \right) - (2 c_1 c_2+N_c c_2^2)\; \cdot \right.
$$
\beq
\left.\left.
\Big( (1+y) T_V^{d'} + \frac{1-y}{3} T_A^{d'} +
\frac{\sqrt{1-4y}}{(1-y)^2} T_V^{s'} +
\frac{(1-4y)^{\frac{3}{2}}}{3(1-y)^2}
T_A^{s'}\Big) \Big] \right.\;
 + \; 2\pi c_D \frac{O_D}{2\pi}\,\right\}\;,
\label{renorm}
\eeq
}
where
\beq
\zeta\;=\;
\left(\frac{\as(m_c)}{\as(m_b)}\right)^{\frac{3N_c}{2\beta_0-4/3}}
\left(\frac{\as(\mu)}{\as(m_c)}\right)^{\frac{3N_c}{2\beta_0}}\;,
\qquad \beta_0\,=\, \frac{11}{3}N_c-2=9 \;.
\label{zeta}
\eeq
We wrote the contribution of the Darwin operator separately although
it is related to the sum $T_V^u+T_V^d+T_V^s$. In this form $c_D$
emerges from the Penguin-type diagrams while the other terms do not
include the annihilation diagrams.

The coefficient $c_D$ of the Darwin operator Eq.~(\ref{dar})
takes the following LLA form:
$$
c_D\;=\; -\frac{1}{2\pi\as(\mu)}\:\zeta\,
\left\{ (1-y)^2 \left(\frac{\eta-1}{n_f} +
\frac{\eta(\xi-1)}{n_f+1}\right)\cdot
\right.
$$
\beq
\left.
\,\left[4 c_1 c_2 -(2 c_1 c_2+ N_c c_2^2)
\left(1+y+\frac{\sqrt{1-4y}}{(1-y)^2}\right) \right] \;-\;
2\frac{\eta(\xi-1)}{n_f+1}\,
\left[2c_1c_2+N_c c_1^2+\frac{n_\ell}{2}\right]\:\right\}\;,
\label{cd}
\eeq
where
\beq
\xi\;=\;
\left(\frac{\as(m_c)}{\as(m_b)}\right)^
{-\frac{2(n_f+1)}{3\beta_0-2}}\;,
\qquad
\eta\;=\;
\left(\frac{\as(\mu)}{\as(m_c)}\right)^{-\frac{2n_f}{3\beta_0}}\;;
\qquad n_f=3\;.
\label{xieta}
\eeq

In principle, there is another source of the Darwin term in the width
which comes from the $1/m_b$ expansion of the expectation value of
$\bar{b}b$ and from the non-logarithmic terms in the expansion of the
transition operator. They were calculated for the case of the semileptonic
width in \cite{bds} and \cite{kapus}.
We can estimate this correction to the LLA neglecting the deviation
of the color factors $c_1$, $c_2$ from their bare values $1$ and $0$,
respectively, and neglecting the mass of the quarks
(leptons) produced by the virtual $W$ boson.
In this approximation the possible effect from the $\bar{u}d$
($\bar{c}s$) loop cancels and we can use the calculations for the
semileptonic widths.
This yields to the leading order in $\as$
\beq
\delta c_D\;\simeq \;
\frac{2N_c+n_\ell}{576\pi^2}
\left(77-88y+24y^2 - 8y^3 - 5y^4 + 36 y^2\ln y \right)\,.
\label{cd2}
\eeq
The overall non-$\log$ term in $c_D$ appears to be of the
opposite sign to the LLA result and is roughly a half of it in
magnitude.  We conclude that the LLA estimate is accurate within a
factor of $2$ being, probably, on the upper side. We use the LLA
expressions for numerical estimates below.

At the next-to-leading order the $\Delta B=1$ weak decay
Wilson coefficients
$c_1(m_b)$, $c_2(m_b)$ are~\footnote{Note that these NLO
values of $c_{1,2}$ are immediately reproduced in the simple LLA if one
uses the more physical $V$-scheme $\as$ coupling \cite{Vscheme}.}
\beq\label{c12}
c_1=1.13\,,\qquad c_2=-0.29\,,
\eeq
corresponding to $\alpha_s(M_Z)=0.118$. This gives
\bea
c_1^2+c_2^2+\frac{2}{N_c}c_1c_2 \;&\simeq 1.15 \;,\qquad\;
2c_1 c_2 + N_c c_2^2 \;&\simeq -0.40 \;,
\nonumber\\
4c_1c_2 \;&\simeq -1.3 \;,\qquad\;
2c_1 c_2 + N_c c_1^2 \;&\simeq 3.2 \;.
\eea
With this input the resulting values
for the coefficients in $\Delta\hat\Gamma$ are
given in Table 2. To illustrate the uncertainty associated with the LLA
we quote two sets of values corresponding to using $\as(m_b)=0.3$ and
to $\as(m_b)=0.2$ (the former option can represent the choice of
the $V$-scheme strong coupling in the LLA expressions, which seems
more appropriate to us).  The coefficients $A$ and $B$ are defined as
$$
\Delta\hat\Gamma = \frac{G_F^2m_b^2}{2\pi}\, |V_{cb}|^2\;
\left(A_V^u O_V^u+ A_V^d O_V^{d'}+ A_V^s O_V^{s'} +
A_A^u O_A^u+ A_A^d O_A^{d'}+ A_A^s O_A^{s'} + \right.
$$
\beq
\left.
B_V^u T_V^u+ B_V^d T_V^{d'}+ B_V^s T_V^{s'} +
B_A^u T_A^u+ B_A^d T_A^{d'}+ B_A^s T_A^{s'} +
2\pi c_D \frac{O_D}{2\pi}\right)
\label{coef}
\eeq
(i.e. using the `redundant' basis including the Darwin operator to show
explicitly the loop contributions).  We also quote the
values of $\tilde B_V^u$, $\tilde B_V^d$ and $\tilde B_V^s$ given by
\beq
\tilde B_V^u = B_V^u - 2\pi\as c_D\,,
\qquad
\tilde B_V^d = B_V^d - 2\pi\as c_D\,,
\qquad
\tilde B_V^s = B_V^s - 2\pi\as c_D\,.
\label{coeft}
\eeq
The Cabibbo mixing is neglected here.

\begin{table}[t]
\begin{center}
\begin{tabular} {|c|c|c|c|c|c|c|c|c|c|c|}
\hline
$\as(m_b)$ & $A_{V,A}^u$ & $A_V^d $ & $A_V^s $ & $A_A^d $  & $ A_A^s $ &
$B_{V,A}^u $ & $B_V^d $  & $B_V^s $ & $B_A^d $ & $B_A^s $  \\
\hline
$0.3$ & $0.98$ & $-0.53$ & $-0.48$ & $-0.15$ & $-0.11$ & $-2.1$ &
$0.70$ & $ 0.64$ & $0.20$ & $0.15$ \\
\hline
$0.2$ & $0.98$ & $-0.53$ & $-0.48$ & $-0.15$ & $-0.11$ & $-2.6$ &
$0.84$ & $ 0.77$ & $0.24$ & $0.18$ \\
\hline
\end{tabular}\vspace*{.5cm}
\begin{tabular} {|c|c|c|c|c|}
\hline
$\as(m_b)$ & $ 2\pi c_D $ & $\tilde B_V^u $ & $\tilde B_V^d $ &
$\tilde B_V^s $ \\
\hline
$0.3$ & $-0.80$ & $-1.3$ & $1.5$ & $1.4$ \\
\hline
$0.2$ & $-0.56$ & $-2.0$ & $1.4$ & $1.3$ \\
\hline\end{tabular}
\end{center}
\caption{Values of Wilson coefficients for the total width
\label{TABLE10}}
\end{table}

For numerical estimates we use the values of the running low-scale
masses $m_b\simeq 4.6\GeV$, $m_c\simeq 1.25\GeV$ and normalize the width
correction $\Delta
\Gamma_{H_b}$ to the semileptonic width which is reliably evaluated in
the OPE (for a review, see \cite{rev}). The final estimates for these
corrections then read as
{\footnotesize
$$
\frac{\Delta\Gamma_{H_b}}{\Gamma_{\rm sl}}  \simeq
\frac{1}{{\rm BR_{sl}}}
\frac{\Delta\Gamma_{H_b}}{\Gamma_{H_b}} \;\simeq\;
0.36\,
\left[
A_V^u \frac{\aver{O_V^u}}{0.02\,{\rm GeV}^3}+ A_V^d
\frac{\aver{O_V^{d'}}}{0.02\,{\rm GeV}^3} +
A_V^s \frac{\aver{O_V^{s'}}}{0.02\,{\rm GeV}^3} \;+ \right.
$$
$$
\left.
A_A^u \frac{\aver{O_A^u}}{0.02\,{\rm GeV}^3}+ A_A^d
\frac{\aver{O_A^{d'}}}{0.02\,{\rm GeV}^3} +
A_A^s \frac{\aver{O_A^{s'}}}{0.02\,{\rm GeV}^3} +
B_V^u \frac{\aver{T_V^{u}}}{0.02\,{\rm GeV}^3} +
B_V^d \frac{\aver{T_V^{d'}}}{0.02\,{\rm GeV}^3}+
B_V^s \frac{\aver{T_V^{s'}}}{0.02\,{\rm GeV}^3} \right.
$$
$$
\left. +\;
B_A^u \frac{\aver{T_A^{u}}}{0.02\,{\rm GeV}^3} +
B_A^d \frac{\aver{T_A^{d'}}}{0.02\,{\rm GeV}^3}+
B_A^s \frac{\aver{T_A^{s'}}}{0.02\,{\rm GeV}^3} +
0.8\, (2\pi c_D) \frac{\rho_D^3}{0.1\,{\rm GeV}^3}
\right]\;=
$$
\vspace*{.2cm}
$$
0.36\: \left[
A_V^u \frac{\aver{O_V^u}}{.02\,{\rm GeV}^3}+ A_V^d
\frac{\aver{O_V^{d'}}}{.02\,{\rm GeV}^3} +
A_V^s \frac{\aver{O_V^{s'}}}{.02\,{\rm GeV}^3} +
A_A^u \frac{\aver{O_V^u}}{.02\,{\rm GeV}^3}+ A_A^d
\frac{\aver{O_V^{d'}}}{.02\,{\rm GeV}^3} +
A_A^s \frac{\aver{O_V^{s'}}}{.02\,{\rm GeV}^3}
\right.
$$
\beq
\left.
+\tilde B_V^u \frac{\aver{T_V^{u}}}{.02\,{\rm GeV}^3} +
\tilde B_V^d \frac{\aver{T_V^{d'}}}{.02\,{\rm GeV}^3}+
\tilde B_V^s \frac{\aver{T_V^{s'}}}{.02\,{\rm GeV}^3}+
B_A^u \frac{\aver{O_V^u}}{.02\,{\rm GeV}^3}+ B_A^d
\frac{\aver{O_V^{d'}}}{.02\,{\rm GeV}^3} +
B_A^s \frac{\aver{O_V^{s'}}}{.02\,{\rm GeV}^3}
\right]\,,
\label{num}
\eeq
}
where $\aver{O_V^u}= \frac{1}{2M_{H_b}} \matel{H_b}{O_V^u}{H_b}\,$, etc.
We recall that the expectation values in $B$ are denoted by $\omega$
for color-straight operators $O$ and by $\tau$ for the octet ones $T$,
Eq.~(\ref{wt}); for $\Lambda_b$ these are $\lambda$ and $-\frac23
\lambda'$, Eq.~(\ref{baryon}).

It is interesting to note that, regarding the
$N_c$ counting rules one can view the Wilson coefficients of the
color-straight operators to be $N_c^0$ while the coefficients of the
octet operators as $1/N_c$. This is true if we recall that formally
$c_1(m_b) = {\cal O}(1)$ while $c_2(m_b) = {\cal O}(1/N_c)$.
These are not
mandatory assumptions for the large-$N_c$ analysis: smallness of a
particular perturbative renormalization can always be compensated by
large logarithms of $M_W/m_b$; in any case the nonleptonic weak decay
coefficients $c_{1,2}$ are external to QCD itself and can be taken
completely arbitrary. Nevertheless, their numerical values fit well
such a naive assignment.

Our procedure of evaluating the $1/m_b^3$ corrections to the
widths then gets justification in the formal $N_c$ counting rules: we
take at face value the $N_c^0$ color-straight expectation values
appearing with the coefficients $\sim N_c^0$, and take only the leading
factorizable values $\sim N_c$ for the color-octet operators which come
with the subleading coefficient $1/N_c$. This formally sums all
leading corrections $\sim N_c^0$ in the decay widths.

\section{Discussion}

We have considered the expectation values of the four-fermion operators
which are encountered in the $1/m_Q$ expansion of the inclusive widths
of beauty hadrons. The size of the color-straight
operators used to be most uncertain in $B$ mesons, since the
factorization approximation a priori is not expected to be
accurate for them. On the other hand, just these operators have the
most direct meaning being analogues of the usual wavefunction density
$|\Psi(0)|^2$. Using the exact relation of their expectation values to
the momentum integral of the elastic transition amplitudes, we
estimated these expectation values employing reasonable
assumptions about the behavior of the formfactors. We showed that the
actual large-$q^2$ asymptotics of the light quark amplitudes in heavy
hadrons is $1/(q^2)^{3/2}$ rather than $1/q^4$ as has been believed
based on simple-minded quark counting rules. We also calculated the
anomalous dimension of the color-straight operators and their
mixing with the octet operators, the effects absent at order $\as$. The
order-$\as^2$ corrections appeared to be numerically enhanced.

In our estimates of the valence expectation values their size obtained
from the two-pole ansatz can be considered as an upper bound. A more
reasonable exponential approximation which suppresses the contributions
of momenta above $1\GeV$, yields smaller results.  We accept it as
a typical lower bound for the color-straight expectation values.
Although the accuracy of the central estimates cannot be too good, they
probably hold better than within a factor of two.

Our estimates, in principle, include a source for
non-valence expectation values. It is related to a different
$q^2$-behavior of formfactors describing different isospin amplitudes
at $q^2<0$. We have it mainly as the different masses
of the isosinglet resonances saturating the formfactors in the
$t$-channel, compared to the corresponding flavor nonsinglet
particles ({\it i.e.}, annihilation shift of masses).
Except for $\eta'$, experimentally these splittings are rather small,
and our literal estimates thus yield a strong suppression. We are
not sure if this really applies to the color-straight expectation
values; the actual suppression can be softer.

We observe a weaker suppression of the non-valence color-straight
matrix elements for the operators with the axial current. It is related
to the nonperturbative `annihilation' effect, in particular, the axial
anomaly in QCD and its solution of the $U(1)$ problem. We conjecture
that the dominant effect is the mass shift of the lowest pseudoscalar
state $\eta'$ while the splitting of the massive resonances (in
particular, axial) or the effect of the possible difference in the
singlet and triplet couplings $G_A(0)$ and $G_A^{(0)}(0)$ is smaller.
Then we get a tentative relation
\beq
\frac{1}{2M_B} \matel{B^+}{\bar{b} \gamma_\mu\gamma_5 b\,
\bar{d} \gamma_\mu\gamma_5 d + \bar{b} \gamma_\mu\gamma_5 b\,
\bar{s} \gamma_\mu\gamma_5 s}{B^+}
\;\approx \;
-\frac{G_A^{(0)}(0)}{8\pi^{3/2}}
\frac{M_{\eta'}^2 M_{\eta'(1295)}^2}{M_{\eta'}+M_{\eta'(1295)}}\;.
\label{nonval}
\eeq
This estimate has the correct scaling $1/N_c$. Numerically, the axial
non-valence expectation values appear to be suppressed by a factor
about $0.1$. We note that the numerical suppressions of various
non-valence effects typically is stronger than the naive factor $1/3$
which can be expected if their justification is merely the large
$N_c=3$.

An interesting indication from our estimates is that the possible
nonperturbative vitiation of the chirality suppression of WA in $B$
mesons emerges at a rather low level (it is governed by the
combinations $\left( \omega_V+\frac{1}{3} \omega_A\right)$,
$\left( \tau_V+\frac{1}{3} \tau_A\right)$). For the color-straight
operators (where the effect a priori can be significant), the literal
suppression is by more than an order of magnitude, in accord with the
evidences from charmed mesons. In our approach the origin of the
suppression roots to the fact that $-G_A(0)\lsim 1/3$. The WA effect of
the octet operators can be probed in the difference of the semileptonic
$b\ra u$ distributions in $B^+$ and $B^0$ \cite{WA}.

The chirality suppression of WA can be eliminated already in the
perturbative evolution of the effective operators. This does not happen
in the LLA \cite{mirage}. Our NLO calculations show that it does not
happen at this level as well. It is interesting to check this property
for the two-loop diagonal renormalization of the color-octet operators.
In any case, we expect it to be lifted in three loops; also, the
non-logarithmic gluon corrections at $k \sim m_b$
defining the initial values of the Wilson coefficients must generate
the chirality non-suppressed effect at some level.

Let us now turn to the phenomenological consequences of our analysis.
The estimated expectation values are typically of the
order of, or somewhat larger than the factorization values (when the
latter are possible) at $\tilde f_B = 160\MeV$ (the factorization value
of $\omega_V$ is additionally
suppressed, and our estimates only partially reproduce this).
The actual expectation values of
the color-straight operators can be smaller if, for example, the
formfactors change sign at $-q^2 \lsim 1\GeV^2$. Such subtleties are
not properly captured by the simple models we relied upon. On the other
hand, larger values than quoted in Table~1 are improbable, at least if
the nonperturbative dynamics we account for are dominated by the momenta
not exceeding $1\GeV$.

The relevance of the latter assumption for the analysis of the
inclusive widths is easy to see, say, on the example of the effect of
interference (dominant in $B$ mesons). The decay rate of the process
$b\ra \bar{u}_{\vec{k}} + \left(cd\right)_q$ is proportional to
$q^2=(p_b-k)^2$. At $k^2=0$ one has $q^2=m_b^2-2p_bk$, and this
constitutes only about $12\GeV^2$ vs. $m_b^2\simeq 21\GeV^2$ already
for $|\vec{k}|=1\GeV$. At the same time the usual relation of the
$1/m_b^3$ effects via the expectation values of the corresponding
four-fermion operators assumes that $q^2=m_b^2$. Therefore, if
$|\vec{k}|$ becomes as large as taken above, the validity of the
leading-order expressions breaks down. In any case, accounting for the
effects like interference in the usual way is legitimate only if their
impact is much smaller than the partonic width of a particular quark
channel. It is worth noting, on the other hand,
that the assumption that the
nonperturbative contributions to the expectation values come from
momenta not exceeding $1\GeV$ is built in the approach of the
QCD sum rules.

At first sight, WA in mesons
and `weak scattering' (WS) in baryons can get
enhanced, in contrast to interference, if the quark momenta saturating
the expectation values of the operators are large. Such a conclusion,
even though eventually may prove to be correct, cannot be justified
{\it a priori}, and even the sign of the corresponding corrections
to the standard expressions is
not known. All such effects manifestly go beyond the $1/m$ expansion
truncated after $1/m_b^3$ terms.  For this reason, simply assuming
large expectation values in $B$ particles does not allow one
to boost significantly the lifetime differences respecting the
self-consistency of the simplest $1/m_b$ expansion.

Bearing in mind all reservations made above, we still quote the central
values for the corrections to the inclusive widths stemming from our
analysis:
$$
\frac{\delta \Gamma_{B^-}}{\Gamma_{\rm sl}}\;\simeq\;
0.36 \left(-1.1_{\rm PI}\, -\,1.2_{\rm D}\right)\;,
\qquad
\frac{\delta \Gamma_{B^0}}{\Gamma_{\rm sl}}\;\simeq\;
0.36 \left(-0.15_{\rm WA}\, -\,1.2_{\rm D}\right)\;,
$$
\beq
\frac{\delta \Gamma_{\Lambda_b}}{\Gamma_{\rm sl}}\;\simeq\;
0.36 \left(2.2_{\rm WS}\,-\,1_{\rm PI}\, -\,1_{\rm D}\right)\;.
\label{finnum}
\eeq
Here we showed separately the effects of different light flavors:
of the operators
$(\bar{b}b)(\bar{u}u)$ responsible for PI in $B$
and WS in $\Lambda_b$, and of $(\bar{b}b)(\bar{d}d)$
generating WA in $B$ and PI in $\Lambda_b$. We singled out the
contribution of the Darwin term. Even though it may seem to be a
computational separation, it is legitimate, for it can be formally
carried through the dependence on the number of light flavors. Being
a flavor singlet, the Darwin operator does not differentiate the
lifetimes of charged and neutral $B$ (also of $B_s$ to the extent that
$SU(3)_{\rm fl}$ is a good symmetry).

The above estimates generally support the original theoretically
predicted pattern of the lifetimes. The non-valence effects seem to be
strongly suppressed. The main effect is destructive PI in $B^-$, about
$-4\%$, while WA is small, at a half percent level.
Moreover, literally we get the
effect of WA decreasing the width, the possibility originally discussed
in \cite{mirage,WA,Ds} and which may seem to contradict the naive
interpretation of WA. The overall difference of $\Gamma(B^-)$ and
$\Gamma(B^0)$ appears about $-4\%$. The major effect is WS in
$\Lambda_b$, $8.5\%$, but it is partially offset by interference,
$-3.5\%$.
The difference between $\Gamma(\Lambda_b)$ and $\Gamma(B^0)$ is
literally $6\%$. These estimates fall close to the expectations
quoted in the review \cite{stone2}. We note that the often discarded
Darwin term (e.g., in \cite{ns2}) typically decreases the width by
about $4\%$, although literally we get its effect in $B$ and
$\Lambda_b$ close to each other. Including it, the overall decrease in
the $\Lambda_b$ lifetime from the four-fermion operators at the order
$1/m_b^3$ comes out only at a percent level while $\tau_{B^0}$
increases by $5\%$ and $\tau_{B^-}$ by $9\%$. The overall
absolute shift is
not too interesting by itself though, since it depends on the
exact definition of the parton width.

It is worth noting that the corrections we addressed do not formally
exhaust the $1/m_b^3$ terms in the asymptotic expansion of
$\Gamma_{H_b}$ -- they come implicitly as well from the expectation
values of the kinetic and chromomagnetic operators which appear at the
level of $1/m_b^2$ corrections. These expectation values in the actual
$b$ hadrons differ from their asymptotic values at $m_b\ra \infty$ by
terms $\sim 1/m_b$ \cite{optical}. In particular, these deviations
contain the expectation value $\rho^3_{LS}$ of one new local
heavy quark operator,
the convection current (or spin-orbital) one. (This operator cannot
appear independently in the expansion of the transition operator
describing the inclusive width since it is not Lorentz-invariant.)
These corrections do not affect $\Gamma_{B^-}-\Gamma_{B^0}$ but, in
principle, are present in $\Gamma_{B}-\Gamma_{\Lambda_b}$.  Their
practical neglection nevertheless is legitimate: such effects are
included in the existing uncertainty of the differences of the
expectation values $\mu_\pi^2$ and $\mu_G^2$ of the $D=5$ operators
between
$B$ and
$\Lambda_b$. So far these expectation values are estimated
without considering corrections to the heavy quark limit; for example,
the value of $\mu_G^2$ in $\Lambda_b$ is nonzero but generally of the
order of $\Lam^3/m_b$.  All such effects are also expected to be
numerically insignificant.  Let us recall that in $B$ mesons the
$\rho^3_{LS}$ expectation value is suppressed to the extent that their
two-particle description is applicable \cite{optical}.

Our analysis does not indicate a crucial impact of the nonfactorizable
contributions in the low-scale expectation values on the $B$ lifetimes
conjectured in \cite{ns2} or later speculations that $\Gamma(B^+)$ can
even exceed $\Gamma(B^0)$ by a significant amount.

The small experimental lifetime of $\Lambda_b$ thus remains a challenge
for the straightforward $1/m_b$ expansion. An accurate measurement of
the semileptonic width of $\Lambda_b$ (or ${\rm BR_{sl}}(\Lambda_b)$)
would help to shed light on the origin of the problem.
The gap between the experimental
value of $\tau_{\Lambda_b}$ and the theoretical expectations could have
been reduced by a significant enhancement of WS and suppression of PI
in $\Lambda_b$, according to the natural guess about
the role of the spectator momentum we mentioned above. Since these
effects originate from the quark decay mode $b\ra c\bar{u}d$
constituting about $60\%$ of the total width, a $25\%$ effect in
the lifetime would signal a more than $50\%$ enhancement of this
channel.  Clearly, such an effect is not possible for the spectator
quark occupying only a small fraction of the total phase space in the
decay, and would require non-conventional composition of the heavy
hadron. The standard calculation of the $1/m_b^3$ terms neglecting the
effect of finite spectator momenta is not applicable for quantitative
description of such large corrections.  For example, the expectation
values of the Darwin operator would be in general much larger, likewise
the mass scale governing the size of higher-dimension operators for
$1/m_b^4 $ and higher-order corrections must be higher in this
situation. \vspace*{0.2cm}

{\it Note added:} When this paper was prepared for publication, a new
improved QCD sum rule calculation of the four-fermion expectation
values appeared \cite{HYC2}; the quoted results correspond to
$v_o\simeq -0.03$, $g_o\simeq 0.003$, and $B_1=0.60\pm 0.01$,
$B_2=0.61\pm 0.01$. It can be suspected, however, that the
stated small errors did not adequately reflect the uncertainties
inherent in the determination from the sum rules {\em per se}.

%*****************
\vspace*{0.2cm}

{\bf Acknowledgments:} \hspace{.4em} D.P. would like to acknowledge
stimulating discussions with B.~Blok on the subject of this paper.
N.U. is grateful to V.~Petrov, P.~Pobylitsa and A.~Vainshtein for
important comments regarding the renormalization of heavy quark
amplitudes, and to I.~Bigi for collaborating on related issues.
This work was supported in part by the NSF under
the grant number PHY~96-05080 and by RFFI grant 96-15-96764;
the work of D.P. was supported by the Ministry of Science and the Arts
of Israel.

\renewcommand{\thesection}{}
\section{Appendices}
\setcounter{section}{0}
\renewcommand{\thesubsection}{A\arabic{subsection}}

\subsection{Combinatorial relations}

\renewcommand{\theequation}{A1.\arabic{equation}}
\setcounter{equation}{0}

Here we quote two general algebraic relations which are useful in
calculating the renormalization of amplitudes containing static heavy
quarks.

For any set of $N$ numbers $x_1,\,...\,, x_N$
\beq
\sum_{k=0}^N \; \left( \Pi_{j=1}^k  \frac{1}{\sum_{l=1}^j -x_l}
\right)
\left(\Pi_{j=k+1}^N  \frac{1}{\sum_{l=j}^N x_l}
\right)\;=\;0
\label{A}
\eeq
(it is assumed that $\Pi_{n_1}^{n_2}=1$ if $n_1>n_2$), and
\beq
\sum_{k=0}^N \; \left( \Pi_{j=k+1}^N  \frac{1}{\sum_{l=k+1}^j -x_l}
\right)
\left(\Pi_{j=1}^k  \frac{1}{\sum_{l=j}^k x_l}
\right)\;=\;0\;.
\label{B}
\eeq
The proof will be given below.

The sums of the type (\ref{A}) are reminiscent to those appearing in
calculating the renormalization of any color-straight operator of the type
$\bar{b}b\cdot O_{\rm light}$. The sums similar to Eq.~(\ref{B}) emerge
in calculations of mixing of an arbitrary heavy quark operator into the
color-straight operators, $\bar{b}\, T \,b \cdot \tilde O_{\rm light}
\ra \bar{b}b\cdot O_{\rm light}$ where $T$ is any color
matrix.

For the color-straight weak vertex $\bar{b} b$ the product of color
matrices on the heavy quark line does not depend on the location of the
weak vertex in respect to the gluon vertices. The $k$-th term in the
sum Eq.~(\ref{A}) corresponds to the diagram where the first $k$ gluons
attach to the initial $b$ quark while the last $N-k$ gluons attach to
the final-state quark, Figs.~4\,a,b. We thus do not sum over
permutations of gluons (their time ordering is fixed) but combine $N+1$
possibilities to place the weak vertex. The analogues of $x_k$ are
$\omega_k$, the energies of gluons flowing into the quark line. With
this identification the structure of the product of the heavy quark
propagators is reproduced.

\thispagestyle{plain}
\begin{figure}[hhh]
 \begin{center}
 \mbox{\epsfig{file=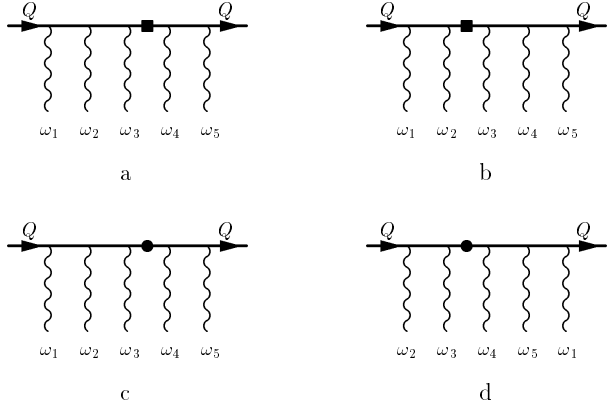,width=8cm}}
 \end{center}
 \caption{The diagrams combined for the color-straight operators (a,b)
and for the mixing into the color-straight
operators (c,d). The solid box denotes a color-straight operator, the
blob in the diagrams c and d stands for an arbitrary heavy quark
operator.  Only two of six ($N=5$) diagrams to be combined are shown in
both cases.}
\label{fig4}
\end{figure}

In dressing a non-straight operator the gluon and weak vertices do not
commute and moving the weak vertex would change the product of color
matrices. However, calculating mixing {\it into} the color-straight
operators amounts to taking trace over color indices of the initial and
final state quarks. Then one, instead, can perform a cyclic move of the
leftmost gluon in the initial state to the latest position in the final
state, and vice versa, Figs.~4\,c,d. In considering Eq.~(\ref{B}) we
thus imply combining all graphs obtained by the cyclic permutations of
a particular diagram. Both considerations apply for any color
representation of the quarks and gluons.

Taken naively, the relation (\ref{A}) would suggest that the
renormalization of the color-straight operators vanishes to all orders
(already in the sum of the above groups of diagrams, before actual
integration over all gluon momenta). Likewise the identity (\ref{B}) would
look like the property that the octet operators never mix with the
straight operators. This is not so, however. The reason is that the
identities Eq.~(\ref{A}) and (\ref{B}) apply only if the external
$b$ quarks are exactly on shell so that their nonrelativistic energy
vanishes, $E=0$. In this case the quark propagators generally become IR
singular when integrated over the gluon momenta, and must be
regularized by a small imaginary part $-i\epsilon$ in each heavy quark
denominator.  Alternatively, this is done
by a shift of the external heavy quark energies by an
infinitesimal imaginary amount. This regularization translates into
the shift of
all $x_l$
which, however, is of the opposite sign for the initial-state and
final-state gluons. For example, for the sum in Eq.~(\ref{A})
$$ x_l \ra x_l+i\epsilon \;\; \mbox{ for $l \le k$ and } \;\; x_l \ra
x_l-i\epsilon \;\; \mbox{ for $l > k$}
$$
(and the opposite shift in
the sum in Eq.~(\ref{B})). This infinitesimal shift of denominators
leads to the fact that the sum of all diagrams does not vanish exactly
but contains certain $\delta$-functions of combination of energies
corresponding to a certain on-shell heavy quark inside the diagrams.
Nevertheless this kills some of the integrations over $\omega$ and
simplifies the remaining integrals.

Let us prove identities (\ref{A}) and (\ref{B}). This can be done most
simply by using the following trick. We can consider the
sum as a rational function of the variable $x_N$ (for example), at
$x_1,\,...\,,x_{N-1}$ arbitrary but fixed. If we show that the residue
of this function at any potential pole vanishes, this would mean that
the whole function vanishes identically.

For the sum in Eq.~(\ref{A}) this is particularly simple. Presence of a
pole means that at certain $k$ some of the denominators with
$j=j_0$ vanish, with either $j_0 \le k$ (to the left of the weak
vertex) or $j_0>k$ (to the right of it). Let $j_0 < k$, for example, and
therefore $\sum_{l=1}^{j_0} -x_l = 0$.
Then the same vanishing denominator will be present for all
diagrams corresponding to $k>j_0$, and it will change only for $k\le
j_0$.  Moreover, all terms with $k>j_0$ will have the common factor $$
\Pi_{j=1}^{j_0-1} \frac{1}{\sum_{l=1}^j -x_l}
$$
which is the product of the propagators to the left of the one which
vanishes.

The remaining factors will be different, but for $k=j_0\!+\!1,
\,...\,,N$ their sum exactly reproduces the l.h.s.\ of Eq.~(\ref{A}) for
the set of $x_{j_0+1}, \,...\,, x_N$ (that is, the case of $N-j_0$
gluons) owing to the on-shellness of the $j_0$-th propagator (the
condition $\sum_{l=1}^{j_0} -x_l = 0$). The induction from the obvious
case $N=1$ immediately proves Eq.~(\ref{A}) for arbitrary $N$.

\thispagestyle{plain}
\begin{figure}[hhh]
 \begin{center}
 \mbox{\epsfig{file=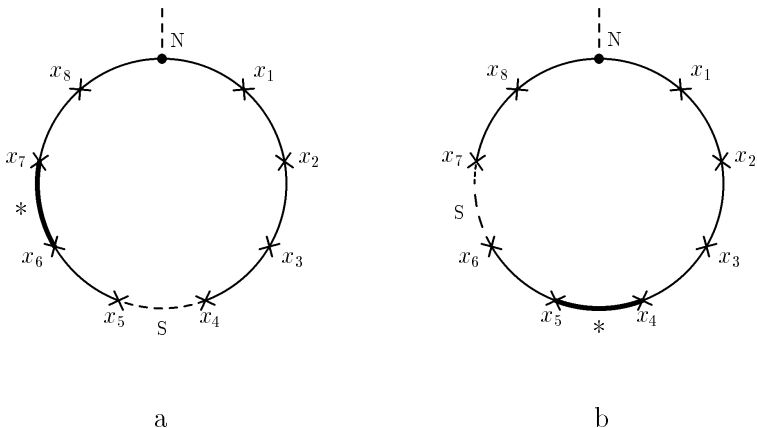,width=10cm}}
 \end{center}
 \caption{Graphic illustration for the case $N=8$.
} \label{fig6} \end{figure}

The proof of the identity Eq.~(\ref{B}) is a little more
complicated. To phrase it, it is convenient to close the heavy quark
line and map it onto the circle, Figs.~5. The weak vertex can be
referred to as North Pole whereas the infinity can be called (with
some reservations) South Pole. Every arc on the circle can be
attributed the corresponding energy denominator. Proceeding from the
$k$-th arc to the $k\!+\!1$-th arc clockwise decreases the
denominator by $x_{k}$.  The values of all denominators are fixed by
the condition that the arc containing the South Pole (the Infinity arc)
has vanishing denominator (correspondingly, it is excluded from the
product of propagators in Eq.~(\ref{B})).

With this image it is easy to establish the vanishing of the residues in
the sum Eq.~(\ref{B}) as well. A pole would appear due to the
vanishing of the denominator of some other arc with $j=j_0$; it
is indicated by the star in Fig.~5a (the Zero arc). This figure shows
the case of $j_0=6$ and $k=4$. It is easy to see that the residue is
exactly canceled by the configuration with $j_0 \leftrightarrow k$,
that is, when the Infinity arc and the Zero arc are interchanged,
Fig.~5b.

Indeed, due to vanishing of the denominators at the both arcs all other
denominators in Fig.~5b are equal to the corresponding denominators in
Fig.~5a. To get the residue one must merely remove the two vanishing
propagators from the product and take it with the factor $-1$ when the
Zero arc is clockwise from the Infinity arc and with the factor
$+1$ otherwise. This cancellation in Eq.~(\ref{B})
reads as
$$ \left(\sum_{l=k+1}^{j_0} -x_l\right)\; \Pi_{j=k+1}^{N}
\frac{1}{\sum_{l=k+1}^j -x_l} \cdot
\Pi_{j=1}^{k} \frac{1}{\sum_{l=j}^k x_l}\;\ra
$$
$$
\left(\sum_{l=k+1}^{j_0} -x_l\right)\; \Pi_{j=j_0+1}^{N}
\frac{1}{\sum_{l=j_0+1}^j -x_l} \cdot \Pi_{j=1}^{j_0}
\frac{1}{\sum_{l=j}^{j_0} x_l} \qquad \mbox{ at } \;
\sum_{l=k+1}^{j_0} -x_l\; \ra\,0\;.
$$

Thus, both identities (\ref{A}) and (\ref{B}) are proved.

\subsection{Two-loop anomalous dimensions}
\renewcommand{\theequation}{A2.\arabic{equation}}
\setcounter{equation}{0}

For the order-$\as$ hybrid renormalization of the heavy quark operators
$\bar{Q} Q \, \bar{q}q$ the identities discussed in Appendix~1 say
that summing over all attachments of the gluon to the heavy quark line
results in $\delta(\omega)$. Therefore, the integration over
${\rm d}^3k$ cannot produce an UV logarithm since it would require an odd
power of $\vec k$ in the integrand. This is not possible in the simple
one-loop diagram. The one loop renormalization of the straight
operators coincides, therefore, with that of the light quark
bilinear, while the octet-to-straight mixing is absent.  For the vector
or axial currents we consider, the overall one-loop renormalization
vanishes.  For the octet operator an additional contribution to
the diagonal renormalization comes from the gluon exchange between the
heavy and light lines.

In order $\as^2$ both the renormalization of the color-straight
operators and the straight--octet mixing occur. We do not consider the
${\cal O}(\as^2)$ diagonal anomalous dimension of the octet operators.
Since the ${\cal O}(\as)$ one does not vanish, the
${\cal O}(\as^2)$ anomalous dimension is scheme-dependent. For the
light-quark currents we are interested in, only
nonfactorizable diagrams must be considered where at least one gluon
connects the heavy quark line with the light part of the diagram.

The hybrid anomalous dimensions are given by a (single) logarithmic UV
divergence of the diagrams in the limit $m_Q\ra \infty$, $|k| \ll m_Q$.
In the Feynman gauge we adopt for computations, only 18 ``double
exchange'' diagrams where two gluons connect light quark line with the
heavy quark line each, yield the $\log$. All other diagrams where there
is only one gluon vertex either on the heavy quark or on the light
quark lines, are finite for symmetry reasons similar to the one-loop
case, or (in the case of dressing the octet operator) yield only the
octet structure we are not interested in.

Combining the diagrams into the groups of three according to the rules
described in
Appendix~1 (all locations of the weak vertex on the heavy quark line
for the color-straight operators, or cyclic permutations of the
$\bar{Q}Qg$ vertices for the octet operators) we get, at fixed values
of the gluon momenta
$k_1$, $k_2$ the sum of the heavy quark propagators in the form
\beq
-2\pi i \delta(\omega_1+\omega_2) \frac{1}{\omega_1+i\epsilon}\qquad
\mbox{ or } \qquad
-2\pi i \delta(\omega_1+\omega_2) \frac{1}{\omega_2+i\epsilon}
\label{a225}
\eeq
for the color-straight operators, or
\beq
-2\pi i\delta(\omega_1+\omega_2) \frac{1}{\omega_1+i\epsilon}
+2\pi i\left( {\cal P}\frac{1}{\omega_1}\delta(\omega_2) -
{\cal P}\frac{1}{\omega_2}\delta(\omega_1) \right)
\label{a232}
\eeq
(and $\omega_1 \leftrightarrow \omega_2$) for the octet operators. In
view of the $\omega \ra -\omega$ symmetry  of the integration only the
structure $-2\pi^2 \delta(\omega_1)\delta(\omega_2)$ survives, and the
resulting integrals contain simple purely three-dimensional expressions
given below . By dimensional counting they all are logarithmic;
they do not vanish since integrations runs over two spacelike vectors.
\vspace*{.3cm}\\
{\bf Dressing of color-straight operators $\bar{Q}Q\, \bar{q}\Gamma
q$} \vspace*{.2cm}\\
The six groups of three diagrams in turn fall into three types which
differ by the location of the gluon vertices on the light quark line,
Figs.~6a-c. Each diagram can have gluon lines twisted or not. Their
expressions are
\bea
& &I_a = \frac{g_s^4}{2} C\,
[\Gamma \gamma_\mu \gamma_0 \gamma_\nu\gamma_0]\,
\int\frac{{\rm d}^3 k_1}{(2\pi)^3} \frac{{\rm d}^3 k_2}{(2\pi)^3}
\frac{(k_1+k_2)_\mu (k_1)_\nu}{\vec k_1^{\,4} \vec{k}_2^{\,2}
(\vec k_1+\vec k_2)^2} \nonumber\\
& &I_b = \frac{g_s^4}{2} C \,
[\gamma_0 \gamma_\mu\Gamma\gamma_\nu\gamma_0]\,
\int\frac{{\rm d}^3 k_1}{(2\pi)^3} \frac{{\rm d}^3 k_2}{(2\pi)^3}
\frac{-(k_2)_\mu (k_1)_\nu}{\vec
k_1^{\,4} \vec{k}_2^{\,4}}
\label{a222}
\\
& &I_c = \frac{g_s^4}{2} C\,
[\gamma_0\gamma_\mu\gamma_0\gamma_\nu\Gamma]
\int\frac{{\rm d}^3 k_1}{(2\pi)^3} \frac{{\rm d}^3 k_2}{(2\pi)^3}
\frac{(k_2)_\mu
(k_1+k_2)_\nu}{\vec k_1^{\,2} \vec k_2^{\,4}
(\vec k_1+\vec k_2)^2}\nonumber\,.
\eea
The color factors $C$ are
\bea\label{two23}
& &C_1 \;=\; [t^a t^b]_l [t^a t^b]_h \;=\;
\frac14\left (1-\frac{1}{N_c^2}\right)\, [1]_l [1]_h \:- \:
\frac{1}{N_c}\, [t^a]_l [t^a]_h\\
\label{two24}
& &C_2 \;=\; [t^a t^b]_l
[t^b t^a]_h \;=\; \frac14 \left(1-\frac{1}{N_c^2}\right) \,[1]_l [1]_h
\;+\; \frac{N_c}{2}\left(1-\frac{2}{N_c^2}\right) \, [t^a]_l [t^a]_h
\qquad
\eea
for ``twisted'' and ``non-twisted'' diagrams, respectively.

\thispagestyle{plain}
\begin{figure}[hhh]
 \begin{center}
 \mbox{\epsfig{file=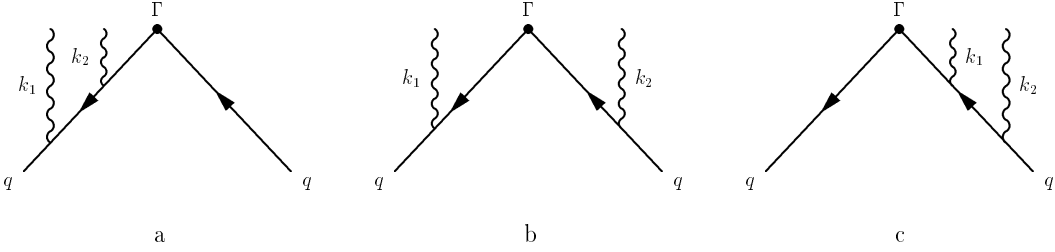,width=14cm}}
 \end{center}
 \caption{Diagrams showing the different attachments of gluons to the
light quark line.}
\label{fig7}
\end{figure}

For $\Gamma=\gamma_0$ or $\gamma_0\gamma_5$, twisted or non-twisted
separately, we have
\beq
I_a+I_b+I_c =
-g_s^4 C\, \Gamma\,
\int\frac{{\rm d}^3 k_1}{(2\pi)^3} \frac{{\rm d}^3 k_2}{(2\pi)^3}
\frac{(\vec k_1\cdot \vec k_2)^2 - \vec k_1^2 \vec k_2^2}
{\vec k_1^{\,4} \vec k_2^{\,4}
(\vec k_1+\vec k_2)^2}
= g_s^4\, C \,\Gamma \;
\frac{1}{32\pi^2}\int^\Lambda\frac{{\rm d}k}{k}\;.
\label{a228}
\eeq
For $\Gamma=\gamma_i$ or $\gamma_i\gamma_5$
$$
I_a+I_b+I_c =
\frac{1}{3} g_s^4 C\, \Gamma\,
\int\frac{{\rm d}^3 k_1}{(2\pi)^3} \frac{{\rm d}^3 k_2}{(2\pi)^3}
\frac{(\vec k_1\cdot \vec k_2)^2
+ 2(\vec k_1\cdot \vec k_2)(\vec k_1^2 + \vec k_2^2)
+ 3 \vec k_1^2 \vec k_2^2}
{\vec k_1^{\,4} \vec k_2^{\,4}
(\vec k_1+\vec k_2)^2}
$$
\beq
=\;  g_s^4\, C \,\Gamma \;
\frac{1}{32\pi^2}\int^\Lambda\frac{{\rm d}k}{k}\;.
\label{a228a}
\eeq
The sum of all diagrams for arbitrary
$\Gamma$ takes the form
\beq
\bar{Q}Q\,\bar{q}\Gamma q\,\ra\,
\left(1+\frac{\as^2}{4}\left(1-\frac{1}{N_c^2}\right)
 \ln{\Lambda}\right)\, \bar{Q}Q\,\bar{q}\Gamma q \;+\;
\frac{\as^2}{4}N_c\left(1-\frac{4}{N_c^2}\right) \ln{\Lambda}\,
\bar{Q}t^a Q\,\bar{q}t^a \Gamma q\,.
\label{a229}
\eeq
\vspace*{.35cm}\\
{\bf Mixing of octet operators $\bar{Q}t^aQ\,
\bar{q}t^a\Gamma q$ into color-straight operators}
\vspace*{.2cm}\\
Taking the trace over the heavy quark color indices we likewise can
combine the 18 diagrams into 6 groups belonging again to the pairs,
where each pair has the same location of the $\bar{q}qg$ vertices
but different trace of color matrices along the heavy line. For
example, for $\Gamma=\gamma_0$ the projection onto the straight operator
yields
$$
\left(I_a+I_b+I_c\right)\vert_{\rm straight} \;=\;
-g_s^4 (C_3+C_4)\, \Gamma\,
\int\frac{{\rm d}^3 k_1}{(2\pi)^3} \frac{{\rm d}^3 k_2}{(2\pi)^3}
\frac{(\vec k_1\cdot \vec k_2)^2 - \vec k_1^2 \vec k_2^2}
{\vec k_1^{\,4} \vec k_2^{\,4}
(\vec k_1+\vec k_2)^2}
$$
\beq
=\; g_s^4 (C_3+C_4)\,\Gamma \;
\frac{1}{32\pi^2}\int^\Lambda\frac{{\rm d}k}{k}\;,
\label{a230}
\eeq
where the color factors
\beq
C_3 = -\frac{1}{4N_c}\left(1-\frac{1}{N_c^2}\right)\,,\qquad
C_4 = \frac{N_c}{8}\left(1-\frac{1}{N_c^2}\right)
\left(1-\frac{2}{N_c^2}\right)\,.
\label{a231}
\eeq
The same renormalization emerges for other Lorentz structures $\Gamma$
as well. \vspace*{.2cm}

For the flavor-singlet operators additional, annihilation
diagrams are possible where the $q\bar{q}$ line forms a closed loop. It
is easy to see that for the vector current it does not contribute. If
the operator is color-straight, only two gluons can come out of the
quark loop. The analogue of the Furry theorem leads to the cancellation
of the two possible diagrams.

For any color-octet operator the sum of
the diagrams where one of the gluons connects the external light and
heavy quark lines yields only the octet operator in analogy with the
one-loop diagrams. All other diagrams can obviously produce only the
octet operators as well.

For the color-straight operator with the axial light quark current,
both gluons must come out of the quark loop. The expression for the
triangle subgraph does not differ from the Abelian case \cite{adler}.
The sum of the diagrams where one of the gluons is attached to the
external light quark and another ends on the heavy quark line yields
only non-logarithmic contribution. The diagrams when both gluons are
absorbed by the light quark legs describe the two-loop
renormalization of the singlet axial current and differ from the
classic Abelian result \cite{adler} only by the color factor  $C_F/2$.
\vspace*{.1cm}

Using Eqs.~(\ref{a229}-\ref{a231}) and the definition
Eq.~(\ref{two66}), we arrive at the ${\cal O}(\as^2)$ matrix of the
anomalous dimensions given in Eq.~(\ref{2loop}) and in the text
following it.

\subsection{Relations between parametrizations}
\renewcommand{\theequation}{A3.\arabic{equation}}
\setcounter{equation}{0}

Here we collect the relations between different parametrizations of the
expectation values of the four-fermion operators in $B$ mesons.

Hadronic parameters suggested in Ref.~\cite{WA} are given by
\bea
\tilde{f}_B^2 M_B\, v_s\;&=&\; -\frac{2}{N_c} \omega_V
\,+\,\frac{2}{3N_c} \omega_A \,-\, 4\tau_V\,+\, \frac{4}{3} \tau_A\\
\tilde{f}_B^2 M_B\, g_s\;&=&\; -\frac{1}{N_c} \omega_V
\,-\,\frac{1}{3N_c} \omega_A \,-\, 2\tau_V\,-\, \frac{2}{3} \tau_A  \\
\tilde{f}_B^2 M_B\, v_o\;&=&\;
- \left(1-\frac{1}{N_c^2}\right) \omega_V
\,+\, \frac{1}{3} \left(1-\frac{1}{N_c^2}\right) \omega_A
\,+\, \frac{2}{N_c} \tau_V\,-\, \frac{2}{3N_c} \tau_A \\
\tilde{f}_B^2 M_B\, g_o\;&=&\;
-\frac{1}{2} \left(1-\frac{1}{N_c^2}\right) \omega_V
\,-\, \frac{1}{6} \left(1-\frac{1}{N_c^2}\right) \omega_A
\,+\, \frac{1}{N_c} \tau_V\,+\, \frac{1}{3N_c} \tau_A\;.
\qquad
\label{apwt2gv}
\eea
The inverse relations read as
\bea
\omega_V \;&=&\;\tilde{f}_B^2 M_B\, \left[-\frac{1}{2} v_o-g_o-
\frac{1}{4N_c}v_s-\frac{1}{2N_c}g_s \right]\\
\omega_A \;&=&\;\tilde{f}_B^2 M_B\, \left[\frac{3}{2} v_o-3g_o+
\frac{3}{4N_c}v_s-\frac{3}{2N_c}g_s \right]\\
\label{a37}
\tau_V \;&=&\;\tilde{f}_B^2 M_B\, \left[\frac{1}{4N_c} v_o+
\frac{1}{2N_c}g_o-
\frac{1}{8}\left(1-\frac{1}{N_c^2}\right)v_s-
\frac{1}{4}\left(1-\frac{1}{N_c^2}\right)g_s \right]\\
\tau_A \;&=&\;\tilde{f}_B^2 M_B\, \left[-\frac{3}{4N_c} v_o+
\frac{3}{2N_c}g_o+
\frac{3}{8}\left(1-\frac{1}{N_c^2}\right)v_s-
\frac{3}{4}\left(1-\frac{1}{N_c^2}\right)g_s \right]\,.
\qquad
\label{gv2wt}
\eea
We recall that for valence quarks $v_s^{\rm fact}=1$ while
$v_o^{\rm fact}=g_s^{\rm fact}=g_o^{\rm fact}=0\,$. \vspace*{.3cm}

For parametrization of \cite{ns2}
\bea
&
\tilde{f}_B^2 M_B\, B_1\;=\; \tilde{f}_B^2 M_B(v_s-4g_s) &=
4(\tau_V+\tau_A) \,+\,\frac{2}{N_c} (\omega_V+\omega_A)
\qquad \qquad \\
&
\tilde{f}_B^2 M_B\, B_2\;=\; \tilde{f}_B^2 M_B(v_s-g_s) &=
-2(\tau_V-\tau_A) \,-\,\frac{1}{N_c} (\omega_V-\omega_A)
\qquad \qquad \\
&
\tilde{f}_B^2 M_B\, \varepsilon_1\;=\; \tilde{f}_B^2 M_B(v_o-4g_o)
&= -\frac{2}{N_c} (\tau_V+\tau_A) \,+\,
\left(1-\frac{1}{N_c^2}\right) (\omega_V+\omega_A)
\qquad \qquad \\
&
\tilde{f}_B^2 M_B\, \varepsilon_2\;=\; \tilde{f}_B^2 M_B(v_o-g_o)
 & =
\frac{1}{N_c} (\tau_V-\tau_A) \,-\,
\frac{1}{2}\left(1-\frac{1}{N_c^2}\right) (\omega_V-\omega_A)\;,
\qquad \qquad
\label{wt2be}
\eea
with the inverse relation
\bea
\omega_V \;&=&\;\tilde{f}_B^2 M_B\,
\left[\frac{1}{4N_c} B_1-
\frac{1}{2N_c} B_2+
\frac{1}{2}\epsilon_1-\epsilon_2 \right]\\
\omega_A \;&=&\;\tilde{f}_B^2 M_B\, \left[\frac{1}{4N_c} B_1+
\frac{1}{2N_c} B_2+
\frac{1}{2}\epsilon_1+\epsilon_2 \right]\\
\tau_V \;&=&\;\tilde{f}_B^2 M_B\, \left[\frac{1}{8}
\left(1-\frac{1}{N_c^2}\right)B_1-
\frac{1}{4} \left(1-\frac{1}{N_c^2}\right)B_2-
\frac{1}{4N_c} \epsilon_1+ \frac{1}{2N_c} \epsilon_2 \right] \\
\tau_A \;&=&\;\tilde{f}_B^2 M_B\, \left[\frac{1}{8}
\left(1-\frac{1}{N_c^2}\right)B_1+
\frac{1}{4} \left(1-\frac{1}{N_c^2}\right)B_2-
\frac{1}{4N_c} \epsilon_1- \frac{1}{2N_c} \epsilon_2 \right] \,.
\qquad
\label{be2wt}
\eea
All these relations hold for each light quark flavor separately.

In the $\Delta B=2$ transitions $B^0_{(s)}\to \bar{B}^0_{(s)}$
determining the width splitting in the $B$-$\bar{B}$ systems one
encounters two four-fermion operators \cite{vsku}, both color-nonsinglet
in the $t$-channel. They are naturally parametrized as
\beq
\matel{B_q}{\bar b_i\gamma_\mu (1-\gamma_5)q^i\:
\bar b_j\gamma_\nu (1-\gamma_5) q^j}{\bar{B}_q}\; =\;
-2\,\tilde{f}_B^2 \left(\tilde v \, P_\mu P_\nu -
\tilde g \, g_{\mu\nu}M_B^2\right)
\label{db2}
\eeq
The non-valence matrix elements vanish.
There is a standard notation $\tilde B_B$ for
$\frac{1}{1+1/N_c} \left(\tilde v-4\tilde g\right)\,$:
\beq
\matel{B_q}{\bar b_i\gamma_\alpha (1-\gamma_5)q^i\:
\bar b_j\gamma_\alpha (1-\gamma_5) q^j}{\bar{B}_q}\; =\;
-2\left(1+\frac{1}{N_c}\right)\,\tilde B_B\,
\tilde{f}_B^2 \,M_B^2\,.
\label{bb}
\eeq
The anomalous dimension of this operator equals two
anomalous dimensions of the $\bar{b} q$ currents, so that $\tilde B_B$
is renorm-invariant in one loop \cite{fb,vslog} (all operators above
are normalized at the low point, not at $m_b$). The combination of the
operators corresponding to the $\tilde v$ structure also renormalizes
multiplicatively in one loop; its anomalous dimension was
calculated in \cite{vsku}. Power mixing of these operators is absent.

\end{document}